\documentclass{aa}
\usepackage[varg]{txfonts}
\usepackage{graphicx}
\usepackage{setspace}
\usepackage{natbib}
\usepackage{longtable}
\usepackage{amssymb}
\bibpunct{(}{)}{;}{a}{}{,}

\begin{document}

\newcommand{\FeII}{[\ion{Fe}{ii}]}
\newcommand{\TiII}{[\ion{Ti}{ii}]}
\newcommand{\SII}{[\ion{S}{ii}]}
\newcommand{\OI}{[\ion{O}{i}]}
\newcommand{\OIp}{\ion{O}{i}}
\newcommand{\PII}{[\ion{P}{ii}]}
\newcommand{\NI}{[\ion{N}{i}]}
\newcommand{\NII}{[\ion{N}{ii}]}
\newcommand{\NIp}{\ion{N}{i}}
\newcommand{\NiII}{[\ion{Ni}{ii}]}
\newcommand{\CaIIp}{\ion{Ca}{ii}}
\newcommand{\PI}{[\ion{P}{i}]}
\newcommand{\CIp}{\ion{C}{i}}
\newcommand{\HeI}{\ion{He}{i}}
\newcommand{\MgIp}{\ion{Mg}{i}}
\newcommand{\MgIIp}{\ion{Mg}{ii}}
\newcommand{\NaI}{\ion{Na}{i}}
\newcommand{\HI}{\ion{H}{i}}
\newcommand{\brg}{Br\,$\gamma$}
\newcommand{\pab}{Pa\,$\beta$}

\newcommand{\macc}{$\dot{M}_{acc}$}
\newcommand{\lacc}{L$_{acc}$}
\newcommand{\lbol}{L$_{bol}$}
\newcommand{\mjet}{$\dot{M}_{out}$}
\newcommand{\mh}{$\dot{M}_{H_2}$}
\newcommand{\Ne}{n$_e$}
\newcommand{\h}{H$_2$}
\newcommand{\kms}{km\,s$^{-1}$}
\newcommand{\um}{$\mu$m}
\newcommand{\lam}{$\lambda$}
\newcommand{\msyr}{M$_{\odot}$\,yr$^{-1}$}
\newcommand{\Av}{A$_V$}
\newcommand{\msun}{M$_{\odot}$}
\newcommand{\lsun}{L$_{\odot}$}
\newcommand{\cm}{cm$^{-3}$}

\newcommand{\bet}{$\beta$}
\newcommand{\alfa}{$\alpha$}

\hyphenation{a-na-ly-sis mo-le-cu-lar pre-vious e-vi-den-ce di-ffe-rent pa-ra-me-ters ex-ten-ding a-vai-la-ble ca-li-bra-tion}

\title{LBT/LUCIFER NIR spectroscopy of PV\,Cephei.}
\subtitle{An outbursting YSO with an asymmetric jet.
\thanks{Based on observations collected at LBT. The LBT 
is an international collaboration among institutions in the United States, Italy and Germany. LBT 
Corporation partners are: LBT Beteiligungsgesellschaft, Germany, representing the Max-Planck Society, 
the Astrophysical Institute Potsdam, and Heidelberg University; The University of Arizona on behalf of the Arizona university system; 
Istituto Nazionale di Astrofisica, Italy;  The Ohio State University, 
and The Research Corporation, on behalf of The University of Notre Dame, University of Minnesota and University of Virginia.}}
\author{A. Caratti o Garatti \inst{1}, R. Garcia Lopez \inst{1},  G. Weigelt \inst{1}, L.V. Tambovtseva \inst{1,2}, V.P. Grinin \inst{1,2,3}, H. Wheelwright \inst{1}, 
\and J.D. Ilee\inst{4,5}}

\offprints{A. Caratti o Garatti, \email{acaratti@mpifr-bonn.mpg.de}}

\institute{
Max-Planck-Institut f\"{u}r Radioastronomie, Auf dem H\"{u}gel 69, D-53121 Bonn, Germany\\
\email{acaratti;rgarcia;weigelt@mpifr-bonn.mpg.de}\\
\and
Pulkovo Astronomical Observatory of the Russian Academy of Sciences, Pulkovskoe shosse 65, 196140 St. Petersburg, Russia\\ 
\email{lvtamb@mail.ru; grinin@gao.spb.ru}\\
\and
The V.V. Sobolev Astronomical Institute of the St. Petersburg University, Petrodvorets, 198904 St. Petersburg, Russia\\ 
\and
School of Physics and Astronomy, EC Stoner Building, University of Leeds, Leeds, LS2 9JT, UK\\
\email{pyjdi@leeds.ac.uk}\\
\and
School of Physics and Astronomy, University of St Andrews, North Haugh, St Andrews, KY16 9SS, UK
   email: jdi3@st-andrews.ac.uk \\
}

%----------------------------------------------------------------------
%
\date{Received date; Accepted date}
%
%----------------------------------------------------------------------
%
%
\abstract
   {Young stellar objects (YSOs) occasionally experience enhanced accretion events, the nature of which is still poorly understood.
   The discovery of various embedded outbursting YSOs has recently questioned the classical definition of EXors and FUors.}
   {We present a detailed spectroscopic investigation of the young eruptive star PV\,Cep, to improve our understanding of
   its nature and characterise its circumstellar environment after its last outburst in 2004.}
   {The analysis of our medium-resolution spectroscopy in the near-IR (0.9--2.35\,$\mu$m), collected in 2012 at the Large Binocular Telescope with the IR spectrograph LUCIFER,
   allows us to infer the main stellar parameters (visual extinction, accretion luminosity, mass accretion and ejection rates), and model the inner disc, jet, and wind.}
  {The NIR spectrum displays several strong emission lines associated with accretion/ejection activity and circumstellar environment.
Our analysis shows that the brightness of PV\,Cep is fading, as well as the mass accretion rate
(2$\times$10$^{-7}$\,M$_{\sun}$\,yr$^{-1}$ in 2012 vs $\sim$5$\times$10$^{-6}$\,M$_{\sun}$\,yr$^{-1}$ in 2004), 
which is more than one order of magnitude lower than in the outburst phase. 

Among the several emission lines, only the [\ion{Fe}{ii}] intensity increased after the outburst. The observed [\ion{Fe}{ii}] emission delineates blue- and red-shifted lobes, 
both with high- and low-velocity components, which trace an asymmetric jet and wind, respectively. 
%Total velocities for the HVC and LVC are -570 and 350\,km\,s$^{-1}$, and -210 and 130\,km\,s$^{-1}$, respectively.
The observed emission in the jet has a dynamical age of 7--8 years, indicating that it was produced during the last outburst. 
The visual extinction decreases moving from the red-shifted ($A_{\rm V}(red)$=10.1$\pm$0.7\,mag) to the blue-shifted lobe ($A_{\rm V}(blue)$=6.5$\pm$0.4\,mag). 
We measure an average electron temperature of 17\,500\,K and electron densities of 30\,000\,cm$^{-3}$ and 15\,000\,cm$^{-3}$ for the blue and the red lobe, 
respectively. The mass ejection rate in both lobes is $\sim$1.5$\times$10$^{-7}$\,M$_{\sun}$\,yr$^{-1}$, approximately matching the high accretion rate observed 
during and immediately after the outburst ($\dot{M}_{out}$/$\dot{M}_{acc}$$\sim$0.05--0.1). The observed jet/outflow asymmetries are consistent with an inhomogeneous medium. 

Our modelling of the CO emission hints at a small-scale gaseous disc ring, extending from $\sim$0.2-0.4\,AU to $\sim$3\,AU from the 
source, with an inner temperature of $\sim$3000\,K. Our \ion{H}{i} lines modelling indicates that most of the observed emission comes from an expanding 
disc wind at $T_{\rm e}$=10\,000\,K. The line profiles are strongly affected by scattering, disc screening, and outflow self-absorption.
}
{According to the classical definition, PV\,Cep is not an EXor object, because it is more massive and younger than typical EXors.
Nevertheless, its spectrum shows the signature of an `EXor-like' outburst, suggesting a common origin.}
\titlerunning{LBT/LUCIFER NIR spectroscopy of PV\,Cep}
\authorrunning{Caratti o Garatti, A. et al.}

\maketitle

%
%------------------------------------------------------------------------- 
%---------------------------------------------------------------------------

\section{Introduction}
\label{introduction:sec}

Young stellar objects (YSOs) are characterised by accretion and ejection processes, which are closely related to and accompany
the whole star-formation period from the protostellar to the pre-main sequence phase, with highly variable strength and duration.

In addition to continuous accretion, YSOs may occasionally experience enhanced accretion events, 
which produce episodic increments of their optical 
and infrared brightness. These objects are called  young eruptive stars, usually divided into FUors (named after FU Orionis, 
the prototype) and EXors (from EX Lupi). FUors exhibit a large brightening up to $\sim$5--6\,mag in the optical lasting from few years
to decades~\citep[][]{H-K}, whereas EXors~\citep[][]{herbig89} display smaller bursts (1--3\,mag) with shorter duration (from days to months), and higher frequency
(every few years). The two subclasses also differ from each other in the observed spectral features in the optical/NIR regime. Namely,
EXors have emission-line spectra, similar to those from Classical TTauri stars~\citep[see e.\,g.,][]{herbig07}, whereas FUors show absorption-line spectra, which render 
distinct spectral types (SpTs) as a function of the wavelength~\citep[i.\,e. F or G SpTs can be inferred in the optical, M SpT in the NIR; see e.\,g.,][]{H-K}.
Moreover, EXors and FUors have been historically classified as low-mass, pre-main sequence objects. However, the discovery of various embedded outbursting YSOs, 
especially in the 
NIR~\citep[e.\,g., \object{OO Ser}, \object{V1647 Ori}, \object{HBC 722}, \object{V2492 Cyg}, \object{V2775\,Ori}; see e.\,g.][]{hodapp96,fedele,kospal,covey,caratti11},
suggests that these processes %may be common in YSOs despite their mass or evolutionary stage,
may take place in YSOs within a wide range of masses and ages %with different mass and at different evolutionary stages, 
and that there might not be a neat separation between the two sub-groups.

These outburst events are likely produced by disc instabilities, and YSOs may accrete a significant ammount of their mass during such 
episodes~\citep[see, e.\,g.,][]{evans09,vorobyov}. 
%although it is not clear which processes originate them~\citep[][{}. 
In addition, enhanced accretion increases the outflow activity~\citep[][]{cabrit90,brittain07}.
The material ejected by such outbursts might produce bright knots along the jet by triggering MHD instabilities in the flow~\citep[see e.\,g.,][]{fendt}.
Like the majority of YSOs, also many young eruptive stars drive outflows, HH or jets. For example, \object{V2492 Cyg} and \object{V346 Nor} are 
known to drive Herbig-Haro objects~\citep[\object{HH 569} and \object{HH 57}, respectively; see][]{bally03,reipurth83}.
On the other hand, \object{OO Ser} is associated with outflow activity and shock-excited H$_2$ emission~\citep[][]{hodapp12}, whereas
\object{Parsamian 21} shows H$\alpha$ knots with dynamical ages of 40--80 years~\citep[][]{staude,kospal08}.
However, it is difficult to associate specific eruptive/outburst events with knots or augmented ejection
along the flows. To the best of our knowledge, \object{Z CMa} is the only case where ejected knots along the flow could be associated 
with observed outbursts~\citep[][]{whelan10}. Thus, the study of such phenomena is of extreme interest, providing us with
information about the accretion/ejection mechanism as a whole.
%ysos may accrete part or most of their mass during these episodic 

%Intro 

Among the outbursting sources, \object{PV\,Cephei} is a unique object.
It is a pre-main sequence young eruptive star, strongly variable, located in between the L\,1155 and L\,1158 clouds. Its distance is debated, 
between 325\,pc~\citep[][]{straizys} and 500\,pc~\citep[][the latter value is assumed throughout the paper]{cohen81},
as well as its spectral type. Although largely recognised as an 
A5~\citep[because of the \ion{H}{i} lines observed in absorption; see e.\,g.,][]{cohen81,the,abraham00},
an absorption spectrum of SpT G8--K0 has also been reported~\citep[][]{magakian01}. The estimated mass of the central source ranges from 2.4\,$M_{\sun}$~\citep{kun}
to $\sim$4\,$M_{\sun}$~\citep{hamidouche} with a massive accretion disc of $\sim$0.8\,$M_{\sun}$~\citep[][]{hamidouche}, which drives a precessing jet 
with Herbig-Haro objects~\citep[HH\,215, HH\,315 and HH\,415;][]{reipurth97} and a massive asymmetric CO outflow~\citep[][]{arce02}.
Adaptive optics (AO) observations exclude the presence of a close companion~\citep[down to $\sim$50\,AU;][]{connelley}, which could trigger the outbursts.

PV\,Cep was spectroscopically and photometrically classified as EXor~\citep[][]{herbig89}, after the outburst in 1977. After exhibiting a large outburst in 
2004, its brightness faded by several magnitudes (up to four in the I and R bands) between 2005 and 2009~\citep{kun}, with a transient peak in 2008.
Its photometric decline was 
caused by the reduced accretion rate and the increased circumstellar extinction~\citep{kun,lorenzetti11}.
Recently, \citet{lorenzetti11} pointed out that PV\,Cep is not a genuine EXor, being more massive and complex than typical solar-mass EXors.

%This source is therefore a unique object, which deserves a deeper investigation.

To clarify the nature of PV\,Cep, we present its medium-resolution spectroscopy in the near-IR (0.9--2.35\,$\mu$m), 
collected at the Large Binocular Telescope with the infrared spectrograph LUCIFER.

This paper is organised as follows.
Section~\ref{observations:sec} illustrates our observations and data reduction.
In Sect.~\ref{results:sec}, we report the results with a detailed description of the detected spectral features, and, in Sect.~\ref{analysis:sec}, 
we derive the main accretion and ejection properties of the YSO and disc physical parameters from  the analysis of its [\ion{Fe}{ii}], \ion{H}{i}, and
CO features.
In Section~\ref{discussion:sec}, we then discuss the accretion/ejection properties in PV\,Cep, its peculiar asymmetric jet and its circumstellar environment.
Finally, our conclusions are drawn in Sect.~\ref{conclusion:sec}.

\section{Observations and data reduction}
\label{observations:sec}

Our spectroscopic observations of PV\,Cep were acquired at the Large Binocular Telescope (LBT) on Mount Graham International Observatory (AZ, USA), 
using the infrared camera and spectrograph LUCIFER at medium resolution \citep{lucifer}. The N1.8 camera was used, corresponding to a spatial scale of 0\farcs25/pixel.
The same spectral dataset was taken on the 23rd and 24th of June 2012 (with an average seeing of 1$\arcsec$ and 0\farcs8 seeing, respectively) and covers the $z$ 
(0.88--1.01\,\um), 
$J$ (1.16--1.31\,\um), $H$ (1.54--1.74\,\um) and $K$ (2.01--2.35\,\um) spectral segments, with a total integration time (per night and per segment) of 1080, 1680, 480 
and 192\,s 
in the $z$, $J$, $H$ and $K$ bands. We adopted the 210\_zJHK grating unit with a 0\farcs5 slit width, leading to a nominal spectral resolution of $\mathcal R \approx$6900, 
8500, 7800 and 6700 in the $z$-, $J$-, $H$- and $K$-bands, respectively. The slit was positioned on the target along the main outer outflow axis (P.A.=348$\degr$).
All the raw data were reduced by using the IRAF\footnote{IRAF (Image Reduction and Analysis Facility) is
distributed by the National Optical Astronomy Observatories, which are operated by AURA, Inc., cooperative agreement with the
National Science Foundation.} package and applying standard procedures. Each observation was flat fielded, sky subtracted, and corrected for the 
curvature derived by long-slit spectroscopy, while atmospheric features were removed by dividing the spectra by a telluric standard star (HD 207636, A0 spectral type), 
normalised to the blackbody function at the stellar temperature, and corrected for its intrinsic photospheric absorption features.
The raw spectra were wavelength calibrated using the bright OH lines detected on each frame~\citep{rousselot}, with average 
uncertainties of 0.2--0.4\,\AA, depending on the number and intensities of the OH lines detected in the considered spectral segment.
The average values of the instrumental profile in the dispersion direction, measured from Gaussian fits to the OH sky lines, 
is 2.1, 2.2, 2.9, 4.9\,\AA, in the $z$, $J$, $H$, and $K$ band segments, respectively. 
Radial velocities of the observed lines were also obtained through single or multiple Gaussian fits in case of blended lines, measured in the observer local standard of rest (LSR) 
and corrected for the cloud speed with respect to the LSR \citep[$v_{cloud}=-3.0$\,km\,s$^{-1}$,][]{torrelles86}.
As a result, the velocity calibration is accurate up to $\sim$10--20\,km\,s$^{-1}$, depending on the considered spectral segment.
Finally, each spectral segment was flux calibrated through a spectrophotometric standard star (Hip 106737) observed using the same instrumental settings 
and slit widths of 0\farcs5 and 1\farcs5, to estimate and correct for the flux losses due to the seeing.

\section{Results}
\label{results:sec}

\begin{figure}
\includegraphics [width=9.5cm]{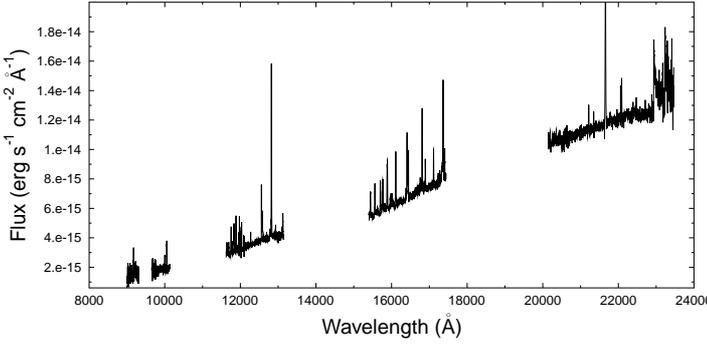}\\
\caption{LUCIFER z-, J-, H- and K-band medium resolution flux-calibrated spectrum of PV\,Cep.}
\label{fig:IRspectrum}
\end{figure}

First, the two spectroscopic datasets were analysed separately to check for variability. There is no clear evidence of variability in both line profiles 
and YSO continuum between the two nights, within the uncertainties ($\Delta$F/F) given by the flux calibration of our spectrophotometric standard 
star (about 20\% in the $I$ band, and between 5\% and 10\% in the $J$, $H$, and $K$ bands)
Therefore, we combined the two datasets to increase the signal to noise ratio of our spectrum.
The final NIR flux-calibrated spectrum of PV\,Cep is presented in Figure~\ref{fig:IRspectrum} and shows a steeply rising continuum with several strong emission lines.
The flux-calibrated spectrum translates into the following $I$, $J$, $H$, and $K$ band magnitudes: 14.7$\pm$0.3, 12.30$\pm$0.08, 10.73$\pm$0.08\,mag,
and 8.80$\pm$0.04, respectively. These values are smaller than those previously observed before, during, and after the outburst in 2004 or the transient peak
in 2008~\citep[see e.\,g.,][]{lorenzetti11,kun}, indicating that PV\,Cep's brightness is slowly fading, especially in the $J$, $H$, and $K$ bands 
($\Delta$m$_I$$\sim$1.5\,mag, $\Delta$m$_J$$\sim$2.1\,mag, $\Delta$m$_H$$\sim$2.6\,mag, and $\Delta$m$_K$$\sim$2.3\,mag, since 2008).

%%%%%%%%%%%%%%%%%%%%%%%%%%%%%%	complete spectrum  %%%%%%%%%%%%%%%%%%%%%%%%%%%%%%%%%%%%%%%%%%%%%%%%%%%%%%%%%%%%%%%%%%%%%%%%%%%
%%%%%%%%%%%%%%%%%%%%%%%%%%%%%%

\subsection{Detected lines}
\label{lines:sec}

%%%%%%%%%%%%%%%%%%%%%%%%%%%%%%	normalised spectrum  %%%%%%%%%%%%%%%%%%%%%%%%%%%%%%%%%%%%%%%%%%%%%%%%%%%%%%%%%%%%%%%%%%%%%%%%%%%
\begin{figure*}
\resizebox{\textwidth}{!}{\includegraphics{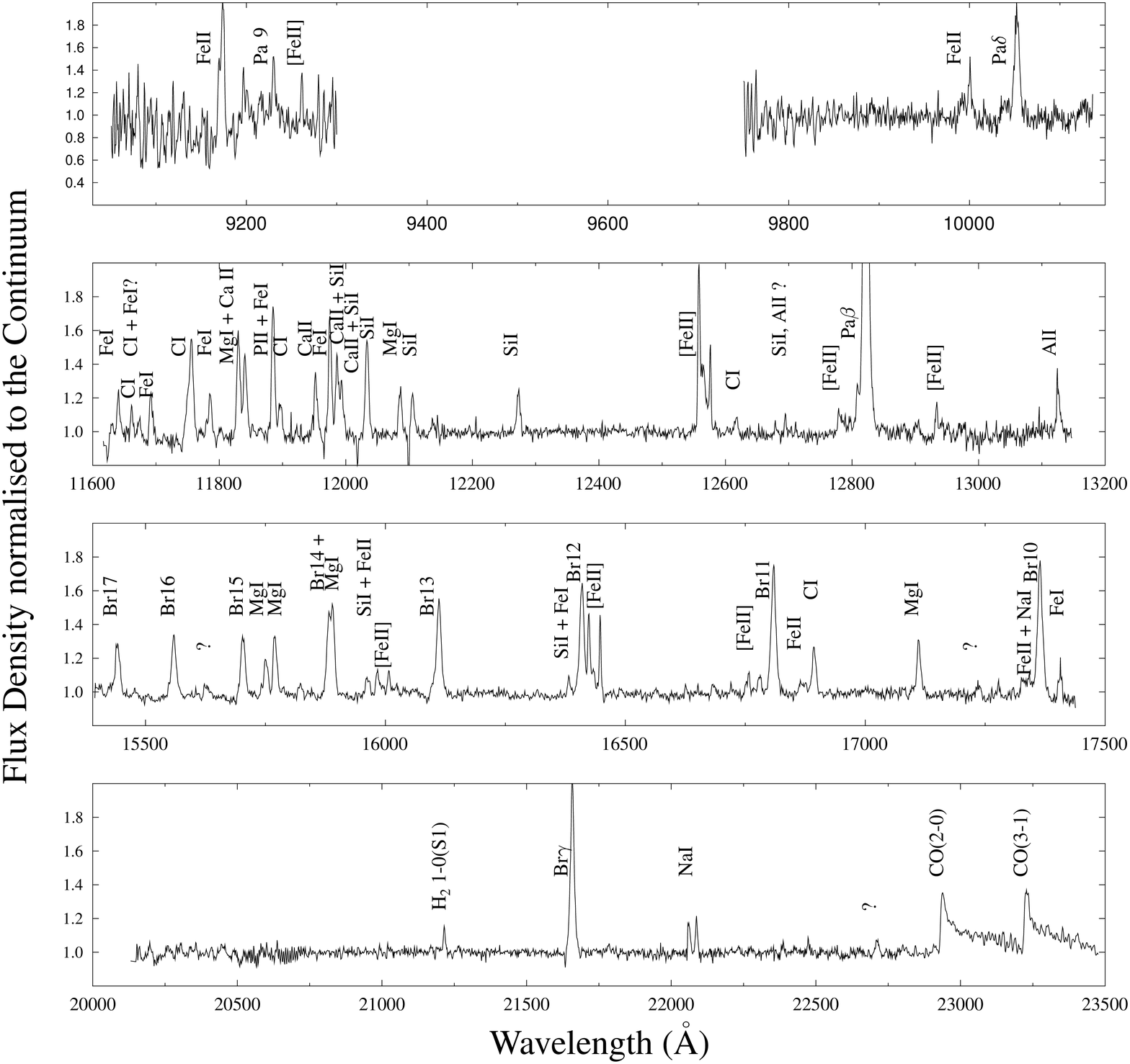}}
\caption{LUCIFER $z-$, $J-$, $H-$ and $K$-band MR spectrum of PV\,Cep normalised to the continuum. The most prominent features are labelled.}
\label{fig:normspectrum}
\end{figure*}
%%%%%%%%%%%%%%%%%%%%%%%%%%%%%%

Figure~\ref{fig:normspectrum} displays our PV\,Cep continuum-normalised spectrum, labelling the most prominent features.
Lines were identified using wavelengths from the NIST Atomic Spectra Database Lines\footnote{http://physics.nist.gov/PhysRefData/ASD/lines\_form.html}.
In addition, NIR spectra from several active Classical TTauri stars (CTTSs) and EXors \citep[e.\,g.,][]{kospal11,rebeca11,caratti12} were also scrutinised to 
recognise the observed features.
We detect several forbidden and permitted lines from atomic and molecular species, but no obvious absorption photospheric feature is visible on the continuum. 
For each detected line, Table~\ref{lines:tab} (shown in Appendix) reports  the observed (vacuum) wavelength ($\lambda_{obs}$), 
the calibrated flux (uncorrected for the extinction) and its uncertainty ($F$,$\Delta F$),
the equivalent width ($W_\lambda$), the full width half maximum ($FWHM$), the observed signal to noise ratio (S/N), the line identification (ID), 
its vacuum wavelength ($\lambda_{vac}$), and the resulting radial velocity ($v_{\rm r}\,(peak)$). The equivalent widths and line fluxes were calculated by 
integrating across the line, after subtracting the continuum. 
A few lines could not be properly identified and are labelled as uncertain (ID + `?') or unidentified (`?') in Tab.~\ref{lines:tab}.

The most prominent lines in the PV\,Cep spectrum are circumstellar features, originating from YSO accretion or inner winds, such as, e.\,g., 
\ion{H}{i}, \ion{Ca}{ii}~\citep[see e.\,g.,][]{muzerolle,natta04,caratti12}, shocks along the jet, such as, e.\,g., [\ion{Fe}{ii}] 
and H$_2$~\citep[see e.\,g.,][]{nisini02,caratti06}, inner disc region (e.\,g.,\ion{Na}{i}, CO)~\citep[see e.\,g.,][]{kospal11,lorenzetti11}, 
or chromospheric activity (e.\,g., \ion{Fe}{i}, \ion{Fe}{ii}, \ion{Mg}{i}, \ion{C}{i})~\citep[see e.\,g.,][]{hamann92-1,hamann92-2,kelly94}.

A few fluorescent emission lines from \ion{Fe}{ii} (i.\,e., at 1.0 and 1.69\,$\mu$m) and \ion{C}{i} (i.\,e., 1.17--1.19\,$\mu$m) have also been detected. They are
usually pumped by UV photons, likely from the Ly$\alpha$ or continuum emission of the stellar photosphere~\citep[see e.\,g.,][]{bautista04,johansson04,lumsden12,walmsley00}.

The detected emission lines are normally observed in the NIR spectra of EXors and very active CTTSs, although the fluorescent lines are more typical of Herbig Ae/Be stars.
Some of the brightest lines, namely \ion{H}{i}, \ion{Mg}{i}, \ion{Na}{i}, CO bandheads, were also detected by \citet{lorenzetti09} during the transient peak in 2008
through NIR low-resolution ($\mathcal R \approx$250) spectroscopy. As for the continuum, absolute line fluxes also show a significant decrease of about 
one order of magnitude,
with the exception of the [\ion{Fe}{ii}] emission lines, which were barely visible in the NIR low-resolution spectrum ($\mathcal R \approx$1200) of \citet[][]{connelley},
taken in July 2007, and below the detection threshold ($\sim$3$\times$10$^{-14}$\,erg\,s$^{-1}$\,cm$^{-1}$) of \citet{lorenzetti09} spectra, taken between 2007 and 2008.

All lines are spectrally resolved, with $FWHM$ values ranging from $\sim$100 to $\sim$250\,km\,s$^{-1}$, showing mostly broad single-peaked profiles.
All Brackett line peaks are systematically blue-shifted by $\sim$$-$20($\pm$15)\,km\,s$^{-1}$, whereas $v_{\rm r}\,(peak)$ of the atomic lines 
(blended and low S/N lines excluded) is around 0\,km\,s$^{-1}$, within a $\pm$15\,km\,s$^{-1}$ uncertainty.
On the other hand, the bright Pa$\beta$ and Pa$\delta$ lines are double peaked, with strong red-shifted (at $\sim$30\,km\,s$^{-1}$) and 
weak blue-shifted (at $\sim$$-$270\,km\,s$^{-1}$) peaks. The [\ion{Fe}{ii}] lines have both high- and low-velocity components (HVC and LVC, respectively) 
in both red- and blue-shifted jet lobes (i.\,e., the bright iron lines show
a peculiar four-peaked profile). Values of the [\ion{Fe}{ii}] peak velocities, averaged over all the detected lines, are $-$265 and $-$99\,km\,s$^{-1}$ 
(blue-shifted HVC and LVC, respectively), and 165 and 50\,km\,s$^{-1}$ (red-shifted HVC and LVC, respectively).
The [\ion{Fe}{ii}] emissions of the HVCs are spatailly extended in our spectral images, indicating that the blue- and red-shifted lobes point northwards and southwards, 
respectively, as also observed in the large-scale CO outflow maps~\citep[][]{arce02}.

\section{Analysis and Physical Properties of PV\,Cep}
\label{analysis:sec}

\subsection{Extinction toward the source and the jet}
\label{reddening:sec}

To obtain an accurate value of the visual extinction ($A_{\rm V}$) towards PV\,Cep stellar photosphere and its circumstellar region, we use 
\ion{H}{i} and [\ion{Fe}{ii}] lines, respectively. The hydrogen atom in the NIR traces regions very close to the central source, whereas
the iron forbidden lines delineate the inner jet from few to several hundred AUs. For each species, we employ transitions arising from the same upper level, 
namely Br$\gamma$ and Pa$\delta$, and [\ion{Fe}{ii}] lines at 1.257 and 1.644\,$\mu$m. We assume that the emission arises from optically thin gas, 
thus their observed ratios depend only on the differential extinction. The theoretical values are derived from the Einstein coefficients and frequencies of the transitions.
[\ion{Fe}{ii}] transition probabilities were taken from \citet{nussbaumer88}. We adopt the \citet{rieke85} extinction law to correct for the differential extinction 
and compute $A_{\rm V}$. From the Br$\gamma$/Pa$\delta$ ratio we infer $A_{\rm V}$=10.8$\pm$0.5\,mag. 
We then calculate the [\ion{Fe}{ii}] 1.25/1.64\,$\mu$m ratio for both HV and LV components in the red- and blue-shifted lobe of the jet, 
obtaining $A_{\rm V}$=10.1$\pm$0.7\,mag and $A_{\rm V}$=10.4$\pm$1.9\,mag (HVC and LVC red), $A_{\rm V}$=6.3$\pm$0.5\,mag and 
$A_{\rm V}$=7.1$\pm$0.9\,mag (HVC and LVC blue). HV and LV extinction values are identical (within the error bar) in each lobe,
thus we derive a weighted mean for the extinction of the red-shifted lobe ($A_{\rm V}(red)$=10.1$\pm$0.7\,mag), and the blue-shifted lobe ($A_{\rm V}(blue)$=6.5$\pm$0.4\,mag).
The difference in the extinction between the red- and blue-shifted lobes is due to the circumstellar matter (mostly caused by the disc in PV\,Cep), and
it is usually observed in protostellar jets close to the source~\citep[see e.\,g.,][]{rebeca08,melnikov}.
Finally, we note that the $A_{\rm V}$ towards PV\,Cep varies with time, due to circumstellar extinction variability,
related to dust condensation in the inner disc region after the outburst and/or to the large amount of dust lifted
by the jet from the disc during and after the outbursts~\citep[][]{kun}. Our values are consistent with those found in literature, ranging between 
9 and 14.5\,mag~\citep[e.\,g.,][]{lorenzetti09,connelley,kun}.

\subsection{Stellar parameters}
\label{stellarparam:sec}

As already mentioned in Sect.~\ref{lines:sec}, our spectrum does not provide any firm conclusion on the spectral type of PV\,Cep. However, 
there are several clues from our data and from the literature indicating that the central object is likely an intermediate-mass young star (2--3\,$M_{\sun}$),
with a relatively early spectral type~\citep[F--A, see also][and references therein]{lorenzetti11}.
In particular, it is worth noting: \textit{a)} the presence of H$_2$O MASER emission~\citep[][]{torrelles86,marvel} with a high isotropic 
luminosity~\citep[$\sim$3$\times$10$^{-6}$\,$L_{\sun}$;][]{torrelles86}, which implies that the exciting source is, at least, an 
intermediate-mass star~\citep[see e.\,g.,][]{anglada}; \textit{b)} the presence of a massive circumstellar disc~\citep[$\sim$0.8\,$M_{\sun}$][]{hamidouche},
uncommonly massive even for a young Herbig Ae star. 

Additionally, there are at least two other relevant indications in our spectrum supporting that the central object is an intermediate-mass young star: 
the presence of UV pumped lines in emission, and the inferred jet velocity. 
It is worth noting that UV pumped lines have been also observed in low-mass eruptive stars. However, they were detected only in the outburst 
phase~\citep[e.\,g. \object{EX Lup}, SpT M; see][]{kospal11}. In the spectrum of PV\,Cep, these features are observed during a lower 
accretion state (likely close to the quiescent phase), supporting the idea that they are excited by the stellar photosphere.
An effective temperature ($T_\alpha$) of 8\,000\,K~\citep[with density $N_H$ of 10$^9$\,cm$^{-3}$;][]{kun} is needed to produce the observed
\ion{Fe}{ii} UV pumped lines~\citep[see e.\,g.,][and references therein]{johansson04,johansson}. Therefore, it is quite likely that $T_\alpha$ is the temperature
of the stellar photosphere ($T_{\rm eff}$), which excites the \ion{Fe}{ii} emission in the circumstellar region close to the disc.

The high jet velocity ($\sim$600\,km\,s$^{-1}$, see Sect.~\ref{jetkinematics:sec}) implies that the central source is massive enough (2--3\,$M_{\sun}$) 
to accelerate the ejecta. For example, assuming that the jet velocity $v_{jet}$ is close to the escape velocity, i.\,e. $v_{jet}\sim(2GM_*/R_*)^{1/2}$, 
and M$_*$=2.6\,$M_{\sun}$, R$_*$ =2.9\,$R_{\sun}$~\citep[][]{the}, we obtain $v_{jet}\sim$580\,km\,s$^{-1}$, very close 
to the value measured in Sect.~\ref{jetkinematics:sec}.

We therefore favour the hypothesis of PV\,Cep being an embedded and young Herbig Ae star~\citep[SpT A5;][]{cohen81}.
The adopted stellar parameters along with their references are summarised in Table~\ref{parameters:tab}.

%%%%%%%%%%%%%%%%%%%%%%%%%%%%%%%%%%%%%%%%%%%%%%%%%%%%%%%%%
\begin{table}[h]
 \caption[]{\label{parameters:tab} PV\,Cephei adopted stellar and disc parameters}
\begin{tabular}{lcc}
 \hline \hline\\[-5pt]
Stellar Parameter &  Value & Reference \\
 \hline\\[-5pt]
Distance   & 500\,pc & 1\\
$M_*$   & 2.6\,M$_{\sun}$ & 2\\
$R_*$   & 2.9\,R$_{\sun}$ & 2\\
$SpT$ & A5  & 1\\
$T_*$ &  8300\,K & 1\\
$M_{disc}$ & 0.8\,M$_{\sun}$ & 3\\ 
$i$ & 62$\degr\pm$4$\degr$  & 3\\
\hline
\end{tabular}
\tablebib{(1)~\citet{cohen81}; (2)~\citet{the}; (3)~\citet{hamidouche} 
}
\end{table}
%%%%%%%%%%%%%%%%%%%%%%%%%%%%%%%%%%%%%%%%%%%%%%%%%%%

\subsection{Accretion properties}
\label{accretion:sec}

Several empirical relationships have been proven effective in deriving YSO accretion luminosity from the dereddened line fluxes of 
different accretion tracers~\citep[see, e.\,g.][]{muzerolle98_BrG,muzerolle01,calvet04,natta06}.
We use Pa$\beta$, Br$\gamma$ line dereddened fluxes (adopting $A_{\rm V}$=10.8\,mag, see Sect.~\ref{reddening:sec}) and the following 
empirical relationships from \citet{calvet00,calvet04},
which have been specifically calibrated for protostars of intermediate mass~\citep[][]{calvet04}:

\begin{equation}
\label{PaB:eq}
Log(L_{acc}/L_{\sun}) = 1.03 \times Log(L_{Pa\beta}/L_{\sun}) + 2.80
\end{equation}

\begin{equation}
\label{BrG:eq}
Log(L_{acc}/L_{\sun}) = 0.90 \times Log(L_{Br\gamma}/L_{\sun}) + 2.90
\end{equation}

As a result, we obtain $L_{acc}(Pa\beta)$=5.8\,$L_{\sun}$, and $L_{acc}(Br\gamma)$=5.6\,$L_{\sun}$. 
For our analysis, we therefore assume an average $L_{acc}$ value of 5.7\,$L_{\sun}$.
We note that using the relationships from \citet{muzerolle98_BrG}, which are calibrated on solar-mass YSOs, we would get larger 
values which are not self-consistent (namely, $L_{acc}$$\sim$8 and 26\,$L_{\sun}$, from the Pa$\beta$ and Br$\gamma$ lines, respectively). 
%This latter consideration provides a further indication that the central source may be an intermediate-mass YSO.
Moreover, it is worth noting that \citet{lorenzetti09} derived $L_{acc}(Pa\beta)$ and $L_{acc}(Br\gamma)$ from their multi-epoch spectroscopy
using \citet{muzerolle98_BrG} relationships. The resulting $L_{acc}(Pa\beta)$ and $L_{acc}(Br\gamma)$ values systematically 
diverge by a factor of three to five (see their Tab.~12), too large even considering the uncertainties on the measured fluxes. 
On the other hand, their data become self-consistent (within the error bars) if Eq.~\ref{PaB:eq} and \ref{BrG:eq} are adopted.
In this case their values are also coincident with $L_{acc}$ estimates of \citet{kun}, ranging from $\sim$80\,$L_{\sun}$, during the outburst in 2004 down to 
$\sim$40\,$L_{\sun}$ in 2010. These values are up to one order of magnitude larger than ours, indicating that $L_{acc}$ of PV\,Cep is still decreasing.

Finally, we get an estimate of the mass accretion rate by combining the obtained accretion luminosity and the adopted stellar parameters.
Given that $L_{acc}$ is the energy released by matter accreting from the co-rotational radius (at $\sim$5\,$R_*$) onto the YSO~\citep{gullbring98}:

\begin{equation}
\label{Macc1:eq}
L_{acc} \sim G M_* \dot{M}_{acc} (1 - R_*/R_{in}) / R_* 
\end{equation}

Therefore, $\dot{M}_{acc}$ is given by:

\begin{equation}
\label{Macc2:eq}
\dot{M}_{acc} = L_{acc} * 1.25 R_* / G M_*
\end{equation}

The resulting value is 2$\times$10$^{-7}$\,M$_{\sun}$\,yr$^{-1}$, i.\,e. at least one order of magnitude lower than what was measured by 
\citet{lorenzetti09} and \citet{kun} during and immediately after the 2004 outburst. This indicates a decrease in the PV\,Cep mass accretion rate since
its maximum in 2004--2005. The current value is, nevertheless, still about one order of magnitude higher than the typical values 
of pre-main sequence (PMS) stars of similar mass~\citep[e.\,g.,][]{calvet04,rebeca06}. This implies that accretion at high rates is still ongoing, as also pointed out by 
the presence of the CO lines in emission~\citep[e\,g.,][]{davis11,biscaya}.
%, and thus that PV\,Cep is younger than its PMS counterparts, as also suggested by its flat SED~\citep[][]{connelley}.

%%%%%%%%%%%%%%%%%%%%%%%%%%%%%%	Fe II PV  %%%%%%%%%%%%%%%%%%%%%%%%%%%%%%%%%%%%%%%%%%%%%%%%%%%%%%%%%%%%%%%%%%%%%%%%%%%
\begin{figure}
\includegraphics [width=9cm]{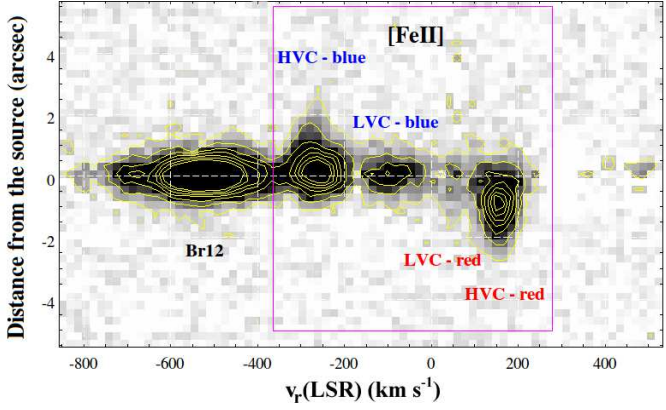}\\
\caption{Position-velocity diagram of PV\,Cep around the 1.644\,$\mu$m [\ion{Fe}{ii}] line in the H-band.
The continuum has been subtracted. The radial velocity on the X-axis refers to the [\ion{Fe}{ii}] line and it is computed in the LSR.
[\ion{Fe}{ii}] blue- and red-shifted high- and low-velocity components as well as the Br\,12 position are marked. 
The distance from the source is indicated on the Y-axis. The YSO position was derived through a gaussian fit to the continuum before
its removal from the spectral image. Positive distance is computed northwards (along the blue-shifted lobe direction).
The contour levels of the spectral image are 3, 10, 20, 30, 40, 50, and 60$\times$ the standard deviation to the mean background.}
\label{fig:PVFeII}
\end{figure}
%%%%%%%%%%%%%%%%%%%%%%%%%%%%%%

%%%%%%%%%%%%%%%%%%%%%%%%%%%%%%	Fe II profile  %%%%%%%%%%%%%%%%%%%%%%%%%%%%%%%%%%%%%%%%%%%%%%%%%%%%%%%%%%%%%%%%%%%%%%%%%%%
\begin{figure}
\includegraphics [width=8cm]{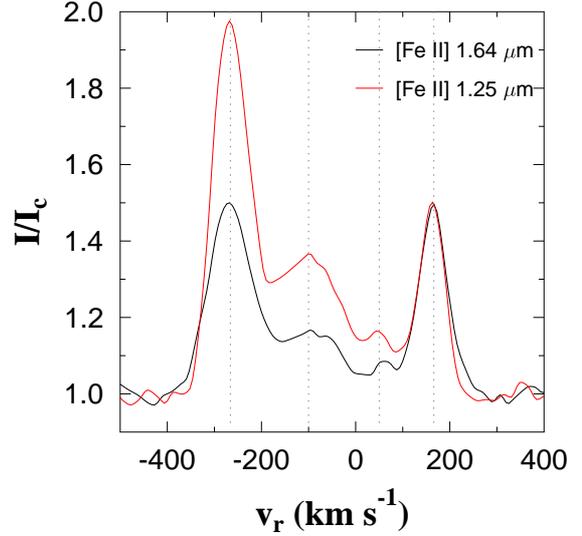}\\
\caption{[\ion{Fe}{ii}] 1.257 and 1.644\,$\mu$m line profiles, normalised to the continuum, plotted on a velocity scale in the LSR. 
The Br\,12 line has been fitted with a Gaussian profile and removed from the 1.644\,$\mu$m line profile.
Average peak velocities at $-$265 and $-$99\,km\,s$^{-1}$ (blue-shifted HVC and LVC, respectively), and 165 and 50\,km\,s$^{-1}$ (red-shifted HVC and LVC, respectively) are marked.}
\label{fig:FeIIprofile}
\end{figure}
%%%%%%%%%%%%%%%%%%%%%%%%%%%%%%

\subsection{Jet kinematics and dynamics}
\label{jetkinematics:sec}

Figure~\ref{fig:PVFeII} shows a continuum-subtracted spectral-image of PV\,Cep around the [\ion{Fe}{ii}] 1.644\,$\mu$m line in the H-band.
Four peaks are detected: two blue-shifted (HVC-blue and LVC-blue) and two red-shifted (HVC-red and LVC-red). Red- and blue-shifted LV components are observed 
only in the brightest lines (at 1.257 and 1.644\,$\mu$m). They are not spatially resolved, observed on source within $\sim$0.8$\arcsec$, or $\sim$400\,AU 
assuming a distance of 500\,pc. On the other hand, the HVC emission extends up to $\sim$2\farcs5 away from the source, tracing part of the jet close to the YSO 
(the wiggling geometry of the jet did not allow us to encompass the jet full extension). The HVC red-shifted emission peaks at $\sim$1$\arcsec$ (or $\sim$500\,AU) 
from the source, whereas we do not detect any obvious knot emission along the blue-shifted part of the jet, although the emission is clearly elongated.
Indeed the actual jet P.A. 
slightly differs from the adopted slit P.A., and our slit encompassed the southern jet lobe better 
than the northern lobe.
This can be clearly seen in Figure~5 of \citet{hamidouche}, where blue- and red-shifted outflow lobes unequivocally show different position angles.
[\ion{Fe}{ii}] profiles of the 1.257 and 1.644\,$\mu$m lines, normalised to the continuum, are shown in Figure~\ref{fig:FeIIprofile}. 
HVC profiles are well defined, with average peak radial velocities of $\sim$$-$265 and 165\,km\,s$^{-1}$ and an average deconvolved line 
width\footnote{$\Delta v$=$\sqrt{FWHM_{line}^2-FWHM_{instrumental}^2}$, where $FWHM_{instrumental} \approx FWHM_{OH}$ in the $H$ band, 
see Sect.~\ref{observations:sec}} ($\Delta v$) of $\sim$70\,km\,s$^{-1}$.
On the other hand, the LVC profiles are less delineated, with average peak radial velocities of $\sim$$-$100 and 50\,km\,s$^{-1}$, and an
average $\Delta v$ of $\sim$160\,km\,s$^{-1}$.
Figure~\ref{fig:FeIIprofile} indicates that both HVC and LVC peak velocities are asymmetric (i.\,e. the blue-shifted peak velocities are larger
than the red-shifted counterparts) by a factor of $\sim$1.6 and $\sim$2, respectively.
Moreover, as already seen in Sect.~\ref{reddening:sec}, the decreasing intensity ratio between the two line profiles implies that the visual extinction increases
from the blue- to the red-shifted lobe.

Interferometric observations of \citet[][]{hamidouche} provide us with a good estimate of the PV\,Cep disc inclination (62$\degr\pm$4$\degr$), indicating
that the system axis is relatively close to the plane of the sky (28$\degr\pm$4$\degr$). From the radial velocity, we can thus derive the tangential and 
total velocity of the jet ($v_{\rm tg}$ and $v_{\rm tot}$, respectively) as well as estimate when the observed knot was ejected. Inferred tangential velocities 
are 500 and 310\,km\,s$^{-1}$ 
for the HV components (blue and red lobe, respectively), and 190 and 90\,km\,s$^{-1}$ for the LV components (blue and red lobe, respectively), 
which translate into $v_{\rm tot}$
of 565, 350, 210, and 110\,km\,s$^{-1}$. Thus, if we assume a $v_{\rm tg}$ of 310\,km\,s$^{-1}$ for the HVC of the red-shifted jet,
we infer that the observed knot has been ejected between 7 and 8 years before our observations (i.\,e. between 2004 and 2005). Therefore, the observed emission 
is likely the outcome of the 2004 outburst~\citep[][]{elek,kun}.

\subsection{Jet physical parameters}
\label{jetphysics:sec}

The various [\ion{Fe}{ii}] lines also allow us to derive the physical properties of the jet close to the source.
These lines come from transitions among the $a^6\!D$, $a^4\!F$, $a^4\!D$, and $a^4\!P$ terms (see Column~7 of Tab.~\ref{lines:tab}). 
Transitions from the first three levels have similar excitation energies ($E_{\rm k} \sim$11\,000--12\,000\,K),
but they have different critical densities ($n_{\rm cr} \sim$10$^4$--10$^5$\,cm$^{-3}$). Therefore, their line ratios can be used to derive the gas electron 
density ($n_{\rm e}$)~\citep[e.\,g.,][]{nisini02,takami}. On the other hand, transitions originating from the $a^4\!P$ term have a higher 
excitation energy ($\sim$20\,000\,K), and they can thus provide an estimate of the electron temperature ($T_{\rm e}$), 
when combined with lines from different energetic levels.

To infer $T_{\rm e}$ and $n_{\rm e}$ of the jet, we use a non-LTE model~\citep[][]{nisini02,rebeca08,rebeca10} 
that considers the first 16 fine-structure levels of [\ion{Fe}{ii}] and thus also includes the lines presented in this paper. 
Our model employs transition probabilities from \citet{nussbaumer88} as well as level energies and rate coefficients 
for electron collisions from \citet{zhang}. 

\begin{figure}
\includegraphics [width=8cm]{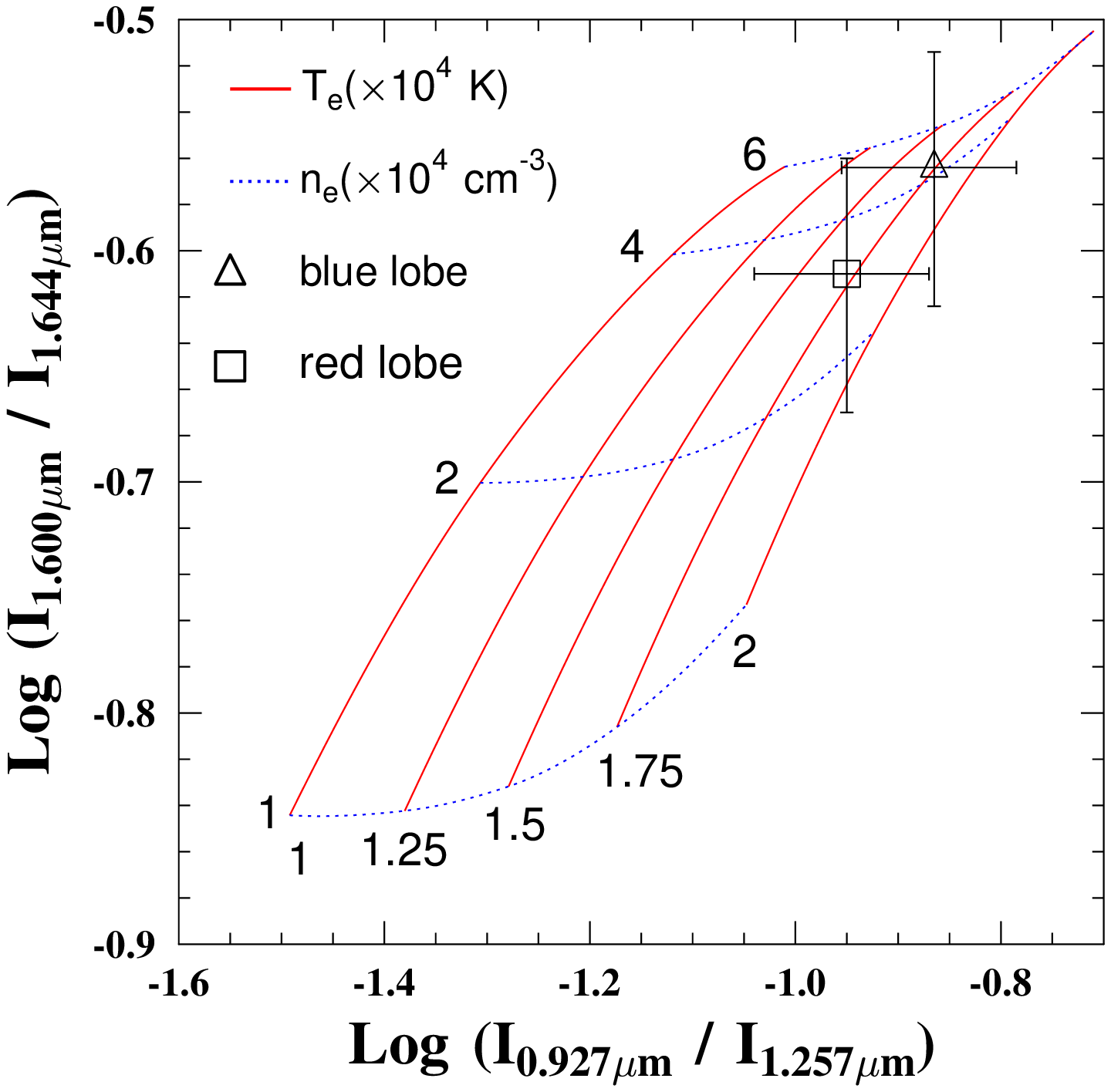}\\
\includegraphics [width=8cm]{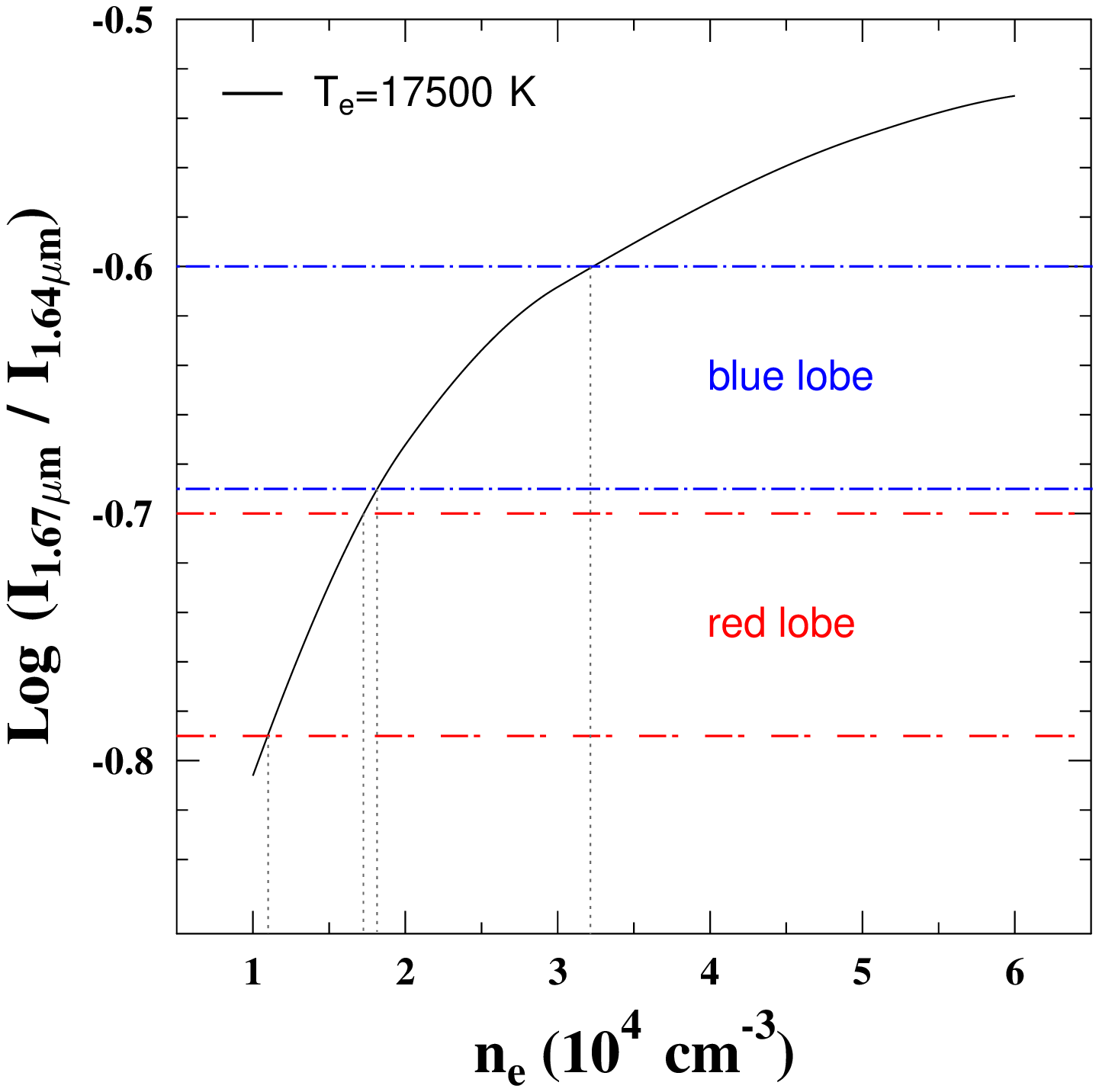}\\
\caption{Electron temperature and density estimates. 
\textbf{Top panel:} Diagnostic diagram employing line ratios sensitive to electron temperature (0.927\,$\mu$m/1.257\,$\mu$m, X axis) and density 
(1.600\,$\mu$m/1.644\,$\mu$m, Y axis).
The grid displays a $T_{\rm e}$ range from 10\,000 to 20\,000\,K (red solid lines), and an $n_{\rm e}$ range from 10\,000 to 60\,000\,cm$^{-3}$ (blue dotted lines).
Triangle and square indicate the logarithmic dereddened line ratios and uncertainties observed in the blue- and red-shifted HVC of the jet, respectively.
\textbf{Bottom panel:} Predicted 1.677\,$\mu$m/1.644\,$\mu$m [\ion{Fe}{ii}] line ratio as a function
of the electron density $n_{\rm e}$. The solid curve corresponds to $T_{\rm e}$=17\,500\,K.
The dashed horizontal lines show the range of ratios observed for the blue and the red lobe.
The intersections between the curve and the dashed lines illustrate the range of $n_{\rm e}$ 
allowed by the ratios observed in the two lobes.}
\label{fig:FeII_Te_ne}
\end{figure}

We first construct a diagnostic diagram (top panel of Figure~\ref{fig:FeII_Te_ne}) using line ratios sensitive to electron temperature (0.927\,$\mu$m/1.257\,$\mu$m, X axis) 
and density (1.600\,$\mu$m/1.644\,$\mu$m, Y axis). The logarithmic theoretical ratios are plotted as a function of $T_{\rm e}$
between 10\,000 and 20\,000\,K (red solid lines) and $n_{\rm e}$ between 10\,000 to 60\,000\,cm$^{-3}$ (blue dotted lines).
We then plot the logarithmic dereddened line ratios observed in the blue- and red-shifted HVC of the jet (triangular and squared symbols, respectively),
using the $A_{\rm V}$ values found in Sect.~\ref{reddening:sec}.
Because we do not detect any red-shifted emission in the 0.927\,$\mu$m line (due to the high visual extinction), we assume the same electron 
temperature for both lobes. However, this assumption is not necessarily correct, because the two lobes have different velocities (Sect.~\ref{jetkinematics:sec}), 
and, possibly, different excitation conditions. The derived $T_{\rm e}$ and $n_{\rm e}$ values for the blue lobe are $17\,500^{+5000}_{-4000}$\,K 
and $45\,000^{+25000}_{-15000}$\,cm$^{-3}$, whereas a slightly lower density value of $30\,000^{+25000}_{-15000}$\,cm$^{-3}$ is inferred for the red lobe.

A further estimate of the $n_{\rm e}$ in both lobes is obtained from the 1.677\,$\mu$m and 1.644\,$\mu$m line ratio (see Figure~\ref{fig:FeII_Te_ne}, bottom panel).
For $T_{\rm e}$=17\,500\,K, we derive $n_{\rm e}(blue)$ between 18\,000 and 32\,000\,cm$^{-3}$ and $n_{\rm e}(red)$ between 11\,000 and 17\,000\,cm$^{-3}$.
Although lower, these estimates are consistent with those previously obtained, confirming that the electron density of the red lobe is between 1.5 and 2 times lower 
than the value obtained in the blue part. Thus we get a weighted mean for the electron density of 30\,000\,cm$^{-3}$ and 15\,000\,cm$^{-3}$ (blue and red lobe, respectively).

%%%%%%%%%%%%%%%%%%%%%%%%%%%%%%	Fe II profile  %%%%%%%%%%%%%%%%%%%%%%%%%%%%%%%%%%%%%%%%%%%%%%%%%%%%%%%%%%%%%%%%%%%%%%%%%%%
%%%%%%%%%%%%%%%%%%%%%%%%%%%%%%

\subsection{Mass ejection rate}
\label{mjet:sec}

Jet physical and kinematical parameters previously inferred permit us to evaluate the mass ejection rate ($\dot{M}_{{\rm jet}}$) in both lobes. 
Moreover, by means of our velocity-resolved observations, we can infer $\dot{M}_{{\rm jet}}$ in the different velocity components, probing 
the structure of the jet/wind itself. 
As the [\ion{Fe}{ii}] emission is optically thin, its luminosity provides us with an estimate of the total mass of the emitting gas ($M_{gas}$).
The mass ejection rate will be then given by $M_{gas} v_{\rm t} / l_{\rm t}$, where $v_{\rm t}$ is the tangential velocity and $l_{\rm t}$ is the jet extension
on the sky plane.
The relation between $M_{gas}$ and [\ion{Fe}{ii}] luminosity is $M_{gas} = \mu m_H (n_H V_{gas}) $, where $\mu = 1.24$ is the average atomic weight, 
$m_{\rm H}$ and $n_{\rm H}$ are the proton mass and the total density, and $V$ is the volume of the emitting region.
The term $(n_H V_{gas})$ can be reformulated as the luminosity of a particular transition (the 1.644\,$\mu$m line, in this case), 
or $n_H V_{gas} = L_{1.644} (h \nu A_i f_i \frac{Fe^{+}}{Fe} \frac{[Fe]}{[H]})^{-1}$,
%\begin{equation}
%\label{method2:eq}
%\dot{M}_k =  \mu m_H n_H \pi r_k^2 v_k   ; \\
%\dot{M}_k = \mu m_H (N_H V) v_t / l_t ; \\
%N_H V = L_X (h \nu A_i f_i \frac{X^{i}}{X} \frac{[X]}{[H]})^{-1}
%\end{equation}
where L$_\mathrm{1.644}$ is the luminosity of the 1.644\,$\mu$m line, for the selected
transition, $A_\mathrm{i}$ and $f_\mathrm{i}$ are the radiative rate and the
fractional population of the upper level of the transition,
$\frac{X^{i}}{X}$ is the ionisation fraction of the considered
species with a total abundance of $\frac{[Fe]}{[H]}$ with respect
to the hydrogen. To obtain the luminosity of the 1.644\,$\mu$m line, we first deredden the fluxes 
of the four observed components, according to the $A_{\rm V}$ values found in Sect.~\ref{reddening:sec}. 
The fractional populations were then computed using a constant value of 17\,500\,K for $T_e$ and $n_{\rm e}$ values of 30\,000\,cm$^{-3}$ 
and 15\,000\,cm$^{-3}$, for the blue and red lobes, respectively (see Sect.~\ref{jetphysics:sec}). We also assume the same $n_{\rm e}$ values
for the LV and HV components. Such assumption implies that both components have similar
physical properties, which is unlikely because they have different spatial extension and might originate from two different circumstellar 
regions (e.\,g., a disc wind and a jet). For example, \citet{rebeca08,rebeca10} find that $n_{\rm e}$ is higher in the LVC of jets
from Class\,I YSOs (about a factor of two with respect to the HVC values), whereas the opposite is found in more 
evolved CTTSs~\citep[see e.\,g.,][]{bacciotti00,coffey08}. In the former case ($n_{\rm e}(LVC) > n_{\rm e}(HVC)$) a lower $\dot{M}_{{\rm jet}}(LVC)$
value (less then a factor of two) is obtained, whereas we get the opposite in the latter case ($n_{\rm e}(LVC) < n_{\rm e}(HVC)$).

Finally, we assume that all iron is ionised (i.\,e. $\frac{Fe^+}{Fe}$=1), with a solar abundance of 2.8$\times$10$^{-5}$~\citep[][i.\,e., no dust depletion]{asplund05}.

As a result, we obtain $\dot{M}_{{\rm jet}}(HVC_{blue})$=1.5$\times$10$^{-7}$\,M$_{\sun}$\,yr$^{-1}$, 
and $\dot{M}_{{\rm jet}}(HVC_{red})$=1.2$\times$10$^{-7}$\,M$_{\sun}$\,yr$^{-1}$
from the high velocity components of the blue and red lobes, respectively, whereas we get 
$\dot{M}_{{\rm jet}}(LVC_{blue})$=5$\times$10$^{-8}$\,M$_{\sun}$\,yr$^{-1}$, and
$\dot{M}_{{\rm jet}}(LVC_{red})$=4$\times$10$^{-8}$\,M$_{\sun}$\,yr$^{-1}$ from the low velocity components of the blue and red lobes. 
Uncertainties can be estimated up to a factor of two. 
Despite the asymmetry in velocity, both lobes have, within the uncertainties, the same mass ejection rates in both HV and LV components, 
whereas the larger $\dot{M}_{{\rm jet}}$ is carried by the HV component, as previously observed in other jets from young and embedded 
stellar objects~\citep[e.\,g.,][]{rebeca08,rebeca10}.

\subsection{CO bandhead modelling}
\label{COmodel:sec}

We investigate whether the CO emission is likely to originate from a small-scale gaseous disc, 
interior to the disc structure already resolved
by \citet{hamidouche}. To evaluate such hypothesis
and constrain the physical properties of the CO emitting region, we
simultaneously model the CO $v=2-0$ and $v=3-1$ bandhead
emission. A detailed description of our
modelling technique is given in \cite{wheelwright} and in
\cite{ilee}. In brief, a geometrically thin flat disc is considered,
where the excitation temperature and CO surface number density
decrease as power laws with increasing radius ($r$) as:

\begin{equation}
\label{Tr:eq}
T({\rm r}) = T_{\rm i} (r / R_{\rm i})^p
\end{equation}

\begin{equation}
\label{Nr:eq}
N({\rm r}) = N_{\rm i} (r / R_{\rm i})^q
\end{equation}

where $T_{\rm i}$ and $N_{\rm i}$ are the excitation temperature and the CO surface density 
at the inner edge of the disc ($R_{\rm i}$).

The optical depth is given by the absorption coefficient per CO
molecule $\times$ the CO column density (i.\,e., $N({\rm r})$ because
of the thin disc). Our model takes into account up to $J$=100
rotational levels in local thermal equilibrium (LTE) for the two
transitions, where the CO energy levels and the Einstein
coefficients are taken from \citet{farrenq} and from \citet{chandra},
respectively. The disc is split into radial and azimuthal cells.
The spectrum of each is calculated individually and then summed together to form the spectrum for the entire disc.

We model the kinematics of the emitting region by assuming Keplerian
rotation ($v_{CO}=\sqrt{GM_*/r}$). We allow the disc to be inclined
and calculate the line of sight velocity for each disc cell. The final
emission feature is given by summation of the individual cell spectra,
which is then convolved to the resolution of the observations. The
best-fit model parameters are determined by comparing the observations
and model using the downhill simplex algorithm implemented in
{\sc{idl}} as the {\sc{amoeba}} routine. We explored several different
initial positions to avoid local minima. The stellar mass and radius
are fixed parameters (as reported in Table~\ref{parameters:tab}), whereas the $T_{\rm i}$, 
$N_{\rm i}$, $i$, inner radius ($R_{\rm i}$), intrinsic line width ($\Delta v$) and the $p$ and $q$
exponents are free parameters. 
Model fits are then related to the data employing the reduced chi-squared statistic. 
The uncertainty in the data is derived from the standard deviation of the flux in the pre-bandhead portion of the spectrum.

\begin{figure}
\includegraphics [width=9.3cm]{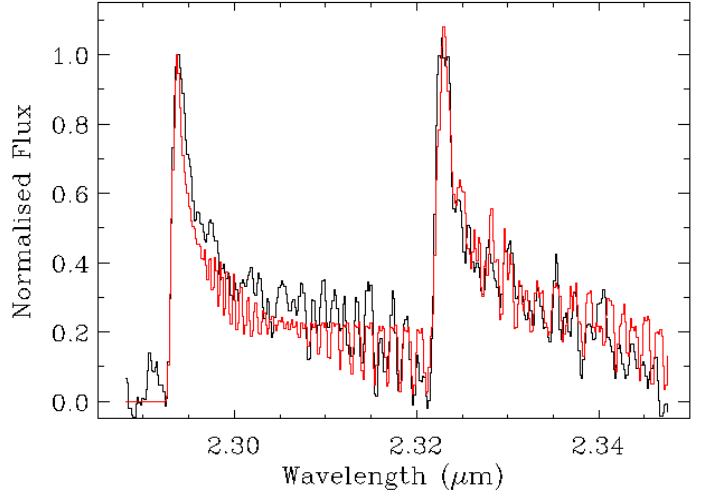}\\
\caption{Close-up around the CO continuum-subtracted spectrum (in black) with the best-fitting model superimposed in red. 
(for a detailed description of the modelling, see Sect.~\ref{COmodel:sec})}
\label{COmodel:fig}
\end{figure}

Figure~\ref{COmodel:fig} shows a close-up of the spectrum around the CO bandheads (in black) with the best-fitting 
model superimposed in red ($\chi_{r}^2$=5.2), whilst the fitted parameters are presented in Table~\ref{COparams:tab}.

%%%%%%%%%%%%%%%%%%%%%%%%%%%%%%%%%%%%%%%%%%%%%%%%%%%%%%%%%
\begin{table}[h]
 \caption[]{\label{COparams:tab} Parameters derived from the best-fit model to the CO bandheads ($\chi_{r}^2$=5.2).
 The outer disc radius ($R_{out}$) is defined at the point in the disc in which the temperature drops below 1000\,K, thus no error is presented.}
\begin{tabular}{lc}
 \hline \hline\\[-5pt]
Physical Parameter &  Value \\
\hline\\[-5pt]
$N_{\rm i}$   & 6$\times$10$^{21\pm1}$\,cm$^{-2}$ \\
$R_{\rm i}$  & 0.3$\pm$0.1\,AU \\
$R_{\rm out}$  & 2.9\,AU \\
$T_{\rm i}$ & 2980$^{+90}_{-640}$\,K \\ 
$i$ & 51$\degr$$^{+25}_{-9}$ \\ 
$\Delta v$ & 7$^{+5}_{-3}$\,km\,s$^{-1}$ \\ 
$p$ &  -0.5$^{+0.4}_{-1.6}$ \\ 
$q$ & -3.9$^{+1.8}_{-1.1}$ \\ 
\hline
\end{tabular}
\end{table}
%%%%%%%%%%%%%%%%%%%%%%%%%%%%%%%%%%%%%%%%%%%%%%%%%%%

%To assess what accretion rates are compatible with the best fitting disc properties, 
%we compare the results of the parameterised disc model with an alpha disc model. We use the analytical formulation described 
%in detail by \citet{Chambers2009}. Essentially, a constant opacity is used to determine the properties of a viscous alpha 
%disc that is flared and irradiated by the central star. In this formulation, the accretion rate is a function of time, 
%which allows us to evaluate the disc properties at a range of accretion rates. To conduct the comparison, we adopt a 
%CO/H$_2$ ratio of 10$^{-4}$ and assume the CO excitation temperature represents the effective temperature of the disc. 
%The stellar properties were set to the values used in the CO fitting procedure and a value of $\alpha =0.1$ was used.

%The properties of the model shown are consistent with a flared viscous disc exhibiting an accretion rate of approx 1e-6 solar masses per year.

%%%%%%%%%%%%%%%%%%%%%%%%%%%%%% CO modelling %%%%%%%%%%%%%%%%%%%%%%%%%%%%%%%%%%%%%%%%%%%%%%%%%%%%%%%%%%%%%%%%%%%%%%%%%%%
%%%%%%%%%%%%%%%%%%%%%%%%%%%%%%

The resulting parameters indicate that the CO bandhead emission can be reproduced using a disc model that has the same inclination 
as the outer disc. Thus the observed CO emission should not originate from the outflow. This supports the existence 
of a small scale gaseous disc structure, that extends from $\sim$0.3$\pm$0.1\,AU up to $\sim$2.9\,AU from the source and has 
an inner excitation temperature 
of $\sim$3000\,K. These quantities are larger than those found in other EXors with lower $M_*$, 
e.\,g., EX\,Lupi~\citep[$R_{\rm i} \sim$0.1\,AU, $T_{\rm CO} \sim$2500\,K, $M_{*} \sim$0.6\,M$_{\sun}$; see,][]{aspin,kospal11}, or
in CTTs ($R_{\rm i} \leq$0.1\,AU, $\Delta R_{CO ring} \sim$0.2\,AU), and more similar
to values found in Herbig Ae stars~\citep[see e.\,g.,][]{salyk}, underlining once more the uncommon EXor nature of PV\,Cep.
Our CO modelling is consistent with a flared irradiated disc, but it is slightly too hot (several hundreds Kelvin). This might be an
aftermath of the recent outburst in 2004, which heated and remodelled the inner disc region.
Finally, assuming a dust sublimation temperature ($T_{\rm sub}$) of 1500\,K~\citep[see e.\,g.,][]{salyk}, from Eq.~\ref{Tr:eq} we can estimate that the sublimation
radius ($R_{\rm sub}$) is positioned at 0.3-0.5\,AU from the source. The same result ($R_{\rm sub}\sim$0.4\,AU) can be obtained from Equation~2 in 
\citet{salyk}\footnote{$T_{\rm sub}=(L_{source}/16 \pi \sigma R_{sub}^2)^{1/4}$} assuming a source luminosity 
($L_* + L_{acc}$) of $\sim$100\,$L_{\odot}$~\citep[see,][and references therein]{lorenzetti09}. The inferred $R_{sub}$ value is very similar to those typical
for Herbig Ae stars~\citep[see e.\,g.,][]{salyk}, and it is comparable to the value inferred in EX\,Lup during the outburst~\citep[$\sim$0.2-0.3\,AU;][]{kospal11}.

\subsection{HI emission modelling}
\label{HImodel:sec}

\begin{figure}
\includegraphics [width=9.3cm]{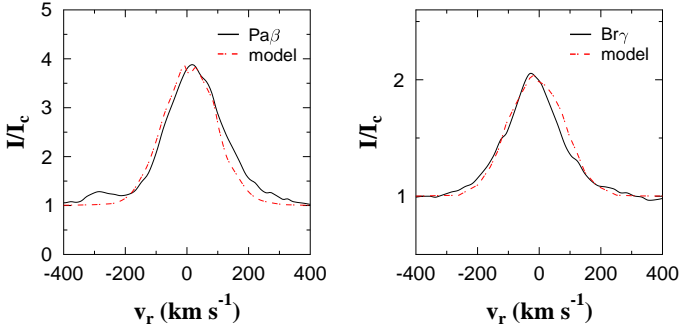}\\
\caption{Pa$\beta$ and Br$\gamma$ observed (in black) and modelled (red dashed lines) profiles (see Sect.~\ref{HImodel:sec}). Lines are normalised to the continuum.}
\label{PaBrprofile:fig}
\end{figure}

As mentioned in Sect.~\ref{results:sec}, the observed lines from the Paschen and Brackett series 
show different shapes and peak velocities: i.\,e., Brackett lines are blue-shifted and single-peaked (at $\sim$$-$20\,km\,s$^{-1}$), 
whereas Paschen lines are double peaked 
(see Figure~\ref{PaBrprofile:fig}), with bright red-shifted maxima (at $\sim$30\,km\,s$^{-1}$) and weak 
blue-shifted peaks (at $\sim$$-$270\,km\,s$^{-1}$).
We aim at modelling both strengths and profiles of the Paschen and Brackett lines to constrain the physical conditions 
and study their origin.

In highly reddened objects, the observed emission is the sum of direct and scattered light, both dimmed
by the circumstellar dust. Depending on the geometry of the scattering and emitting regions,
the line profile of the scattered radiation can strongly differ from the profile given by the direct radiation~\citep[][]{grinin12}. 
This circumstance has to be taken into account when analysing emission spectra of
objects like PV\,Cep. Moreover, extinction %($A_{\rm V}$$\sim$10\,mag, see Sect.\ref{reddening:sec}) 
differentially affects Paschen and Brackett lines observed in the infrared region of
the spectrum. %(producing, e.\,g., an attenuation of $\sim$13 times for the Pa$\beta$, and $\sim$3 for the Br$\gamma$ line). 
Finally, scattering on dust also takes place in a different way for lines of the two considered series 
(e.\,g., the scattering coefficient for the Pa$\beta$ line is a factor of 2.6 greater than the Br$\gamma$ line one).

To model the \ion{H}{i} lines, we therefore consider direct radiation, originating from both 
magnetospheric accretion and disc wind, as well as emission scattered by the circumstellar disc. 
Additionally, our model also takes into account a screening effect from the disc. 
An opaque disc can be a screen that shifts the emission lines towards
the blue part of the spectrum. We assume that the disc is opaque beyond 
the sublimation radius ($\sim$30\,$R_*$ for PV\,Cep). 
Therefore, the remote red-shifted part of the wind is partially hidden from the observer. 
A detailed description of the disc wind, scattering, and magnetospheric accretion models employed
here can be found in \citet{grinin11}~\citep[also see][]{weigelt}, 
\citet{grinin12}, and \citet{tambovtseva}, respectively, whilst a complete description 
of the full model, which groups the aforementioned models, is given in Tambovtseva et al. (in preparation). 
For our modelling, we consider an inclination of 60$\degr$ (angle between the viewing direction and the system axis), as derived from the observations (see Tab.~\ref{parameters:tab}, 
and Sect.~\ref{COmodel:sec}). We let vary the mass accretion and ejection rate values one order of magnitude 
around those found in Sections~\ref{accretion:sec} and ~\ref{mjet:sec}.
%First, to derive the physical conditions of the gas, we construct an excitation diagram for the \ion{H}{i} Brackett series (see, Fig.~\ref{Brdecrement:fig}).
%Line intensities were dereddened ($A_{\rm V}$=10.6\,mag, see Sect.~\ref{reddening:sec}), and their ratios with respect to the Br$\gamma$ 
%are shown in figure as a function of their upper quantum number ($N_{up}$). Superimposed to the data, we also show the best fit
%obtained from our modelling. {\bf PUT BEST FIT values HERE}.

%%%%%%%%%%%%%%%%%%%%%%%%%%%%%% Br decrement %%%%%%%%%%%%%%%%%%%%%%%%%%%%%%%%%%%%%%%%%%%%%%%%%%%%%%%%%%%%%%%%%%%%%%%%%%%
%\begin{figure}
%\includegraphics [width=9.3cm]{br.eps}\\
%\caption{Ratio of the PV\,Cep dereddened Brackett lines with respect to Br$\gamma$ plotted as a function of the upper quantum number ($N_{up}$; filled black circles).
%Superimposed, our best fit model, assuming an expanding wind in LTE at $T_{\rm e}$=10\,000\,K.}
%\label{Brdecrement:fig}
%\end{figure}
%%%%%%%%%%%%%%%%%%%%%%%%%%%%%%
%We calculated several models of
%the disc wind for the different inclinations and tried to compose
%The computed synthetic spectrum consists of the attenuated directPV\,Cep
%and scattered radiation. Figures 1 and 2 presents the result of
%calculations. 
The main parameters of our disc wind model are the following~\citep[see also Fig.~A.1 in][for a graphical representation of the disc wind model 
along with the listed parameters]{weigelt}. 
The disc wind launching region ($\omega$) extends from 3 to 37\,$R_*$ (0.04--0.5\,AU); half-opening angles ($\theta$) of the disc wind for the first and last streamlines are 
30$\degr$--80$\degr$; the parameter
$\gamma$ ($\dot{M}_{wind}(\omega)=\omega^{-\gamma}$, where $\omega=l sin \theta$ is the distance of the point $l$,$\theta$ from the rotation axis), 
which `distributes' the mass load among the streamlines, is 5. The electron temperature is constant and equal to 10\,000\,K, 
the mass loss rate is 3$\times$10$^{-8}$\,M$_\odot$\,yr$^{-1}$, 
and $\beta$=4, where the parameter $\beta$ is a power index in the velocity law 
($v(l)=v_0 + (v_{\infty} - v_0)(1-l_{\rm i}/l)^\beta$, where $v_0$ and $v_{\infty}$ are the initial and terminal values of the radial velocity). 
A scaling factor ($f$) between the radial and tangential velocities decreases with the distance from the star, 
due the magnetic field intensity decrease along the disc. 
%The radius of the emitting region considered when integrating is
%200\,$R_*$ ($\sim$2.7\,AU).
%We note, however, that most of the Brackett emission in our modelling originates from a
%region close to the base of the disc wind (i.\,e. to the disc
%surface), whereas the Paschen emitting region appears more extended.
Table~\ref{HIparams:tab} lists the main parameters of our best-fit model.
%the resulting sizes of the Paschen and Brackett emitting regions are different: $\sim$200\,$R_*$ ($\sim$2.7\,AU),
%and $\sim$10\,$R_*$ ($\sim$0.14\,AU), respectively.
%These two figures show how strongly the shape and intensity of the
%line profiles change depending the mass loss rate. More
%substantial changes in the profiles take place if the mass loss
%rate is high enough (Fig. 4). Then the role of the screen
%increases. Note that the Pa$\beta$ and Br$\gamma$ lines vary in a
%different way under the same conditions. It is due to the
%intrinsic properties of the lines as mentioned above.
In Figure~\ref{PaBrprofile:fig}, we show the Pa$\beta$ and Br$\gamma$ profiles (red dashed lines) resulting from our modelling superimposed 
over the observed profiles (black solid lines).

%%%%%%%%%%%%%%%%%%%%%%%%%%%%%%%%%%%%%%%%%%%%%%%%%%%%%%%%%
\begin{table}[h]
 \caption[]{\label{HIparams:tab} Parameters of the best-fit model to the Pa$\beta$ and Br$\gamma$ lines.}
\begin{tabular}{lc}
 \hline \hline\\[-5pt]
Physical Parameter &  Value \\
\hline\\[-5pt]
$\omega$   & 3--37\,$R_*$ (0.04--0.5\,AU)\\
$\theta$  & 30$\degr$--80$\degr$ \\
$\gamma$  & 5 \\
$\beta$  & 4 \\
$T_{\rm e}$ & 10\,000\,K \\ 
$i$ & 60$\degr$ \\ 
$\dot{M}_{wind}$ & 3$\times$10$^{-8}$\,M$_\odot$\,yr$^{-1}$ \\ 
\hline
\end{tabular}
\tablefoot{In our modelling, $T_{\rm e}$ and $i$ are fixed parameters, whereas the remaining parameters are variable.}
\end{table}
%%%%%%%%%%%%%%%%%%%%%%%%%%%%%%%%%%%%%%%%%%%%%%%%%%%

%The magnetospheric emission is assumed to proceed from a compact, flat and rotating region.
%The main parameters of the magnetospheric model~\citep[see][]{tambovtseva} are: 
%corotation radius ($R_{\rm c}$) equal to 3\,$R_*$, rotation velocity ($v_{\rm rot}$) equal to 100\,km\,s$^{-1}$, a costant radial
%velocity ($v_{\rm r}$) of 15\,km\,s$^{-1}$,
%half-height ($h$) equal to 0.75\,$R_*$, $T_{\rm e} = T_0(r/R_*)^{\alpha_T}$, where $T_0$=8\,000\,K and ${\alpha_T}$=-3,
%and $\dot{M}_{acc}$=10$^{-7}$\,M$_\odot$\,yr$^{-1}$.

%%%%%%%%%%%%%%%%%%%%%%%%%%%%%%%%%%%%%%%%%%%%%%%%%%%%%%%%%
%\begin{table}[h]
% \caption[]{\label{HIparams:tab} Parameters of the best-fit model to the Pa$\beta$ and Br$\gamma$ lines.}
%\begin{tabular}{lclc}
% \hline \hline\\[-5pt]
%Parameter &  Value & Parameter &  Value\\
%\hline\\[-5pt]
%$\omega$   & 3--37\,$R_*$ & $R_{\rm c}$ & 3\,$R_*$ \\
%$\theta$  & 30$\degr$--80$\degr$ & $v_{\rm r}$ & 15\,km\,s$^{-1}$ \\
%$\gamma$  & 5 & $h$ & 0.75\,$R_*$ \\
%$\beta$  & 4 & $T_0$ & 8\,000\,K \\
%$T_{\rm e}$ & 10\,000\,K & ${\alpha_T}$ & -3 \\ 
%$i$ & 60$\degr$ & $\dot{M}_{acc}$ & 10$^{-7}$\,M$_\odot$\,yr$^{-1}$\\ 
%$\dot{M}_{wind}$ & 3$\times$10$^{-8}$\,M$_\odot$\,yr$^{-1}$  & & \\ 
%\hline
%\end{tabular}
%\end{table}
%%%%%%%%%%%%%%%%%%%%%%%%%%%%%%%%%%%%%%%%%%%%%%%%%%%

We then also check for the influence of the magnetosphere on the line profiles applying the magnetospheric model from ~\citet{tambovtseva},
and using an $\dot{M}_{out}$/$\dot{M}_{acc}$ ratio ranging from 0.1 to 0.3.

From the line profile fitting we note that the contribution of magnetospheric accretion to the line profile is small in comparison to the contribution of the disc
wind. In our modelling magnetospheric accretion slightly modifies the width of the line, which is increased less then 10\%
of its FWHM. Therefore, it is not possible to derive a precise estimate of the mass accretion rate from the line modelling. 
On the other hand, both screening effect and the scattering shift the line profiles 
towards the blue side. It is worth noting that both Brackett and Paschen lines would be then blue-shifted, but the shift of the Pa$\beta$ line is partly compensated 
by the self-absorption effect from the outflow, because the optical depth of this line is larger than that of the Br$\gamma$ line.

Finally, the additional blue-shifted peaks in the Paschen lines could not be reproduced by our model. 
Possibly, this weak emission originates from the jet, which is not included in our model.
Indeed, \citet{whelan04} observed extended Paschen emission associated to jet tracers in several CTTs.
The velocity peaks of the secondary maxima in our Paschen lines ($\sim$$-$270\,km\,s$^{-1}$) have the 
same values of the [\ion{Fe}{ii}] blue-shifted HVC.
Nonetheless, no obvious red-shifted component is visible in the Paschen profiles. It might be overwhelmed by the main emission,
assuming that the red-shifted emission of the jet is also asymmetric. This would explain the faint bump observed on the red-shifted side of 
the line (see Figure~\ref{PaBrprofile:fig}, left panel).
Otherwise, its non-detection might be due to the
opaque disc, which blocks the emission from the counter jet. In this case, the Paschen emission from the jet should be very close to the source. 
Assuming that the disc is opaque up to 100\,AU,
it should be closer than $\sim$60\,AU (assuming an inclination angle of 30$\degr$ with respect to the plane of the sky).

%%%%%%%%%%%%%%%%%%%%%%%%%%%%%% Br model %%%%%%%%%%%%%%%%%%%%%%%%%%%%%%%%%%%%%%%%%%%%%%%%%%%%%%%%%%%%%%%%%%%%%%%%%%%
%%%%%%%%%%%%%%%%%%%%%%%%%%%%%%

%%%%%%%%%%%%%%%%%%%%%%%%%%%%%%	HI profile  %%%%%%%%%%%%%%%%%%%%%%%%%%%%%%%%%%%%%%%%%%%%%%%%%%%%%%%%%%%%%%%%%%%%%%%%%%%
%\begin{figure}
%\includegraphics [width=8cm]{HI_norm2.eps}\\
%\caption{Br and Pa lines profiles.}
%\label{fig:HI}
%\end{figure}
%%%%%%%%%%%%%%%%%%%%%%%%%%%%%%%%%%%%%%%%%%%%%%%%%%%%

\section{Discussion}
\label{discussion:sec}

%\subsection{Comparison with literature data}

\subsection{The accretion/ejection connection}
\label{acc-ej:sec}

The detection of an emerging knot, which can be clearly linked to the PV\,Cep outburst event in 2004, is indeed an
interesting result, which supports the hypothesis that enhanced accretion also generates augmented ejection along YSO outflows.
Boosted winds after outbursts were already observed in \object{Z CMa}~\citep[][]{benisty10} , \object{V1647 Ori}~\citep[][]{brittain}, and 
\object{EX Lup}~\citep[][]{goto,kospal11} by means of \ion{H}{i} and CO emission lines. Their intensity has been found to be closely related with
the accretion rate variability. However, it is extremely rare that knot formation is observed along the jet immediately after an outburst 
event from an EXor, because EXors usually do not have jets nor show jet line tracers.
\citet{whelan10} report the first case observed in the Z\,CMa companion, which displayed four `EXor-like' outbursts in the past twenty years.
The authors present AO imaging in [\ion{Fe}{ii}], showing four knots, whose dynamical ages are compatible with the timing of the outbursts.
Thus our data strengthen the idea that knots along the flows may have
been produced by episodic accretion, possibly similar to the `EXor-like' outbursts, i.\,e. bursts of short duration, and not as powerful as those 
from FUors~\citep[][]{ioannidis}. In principle, FUor bursts might be so powerful that they disrupt the circumstellar environment and the small-scale magnetic field. 
Thus the collimated jet/outflow would temporarily switch to a wide-angled wind, failing in developing knots along the flow. This might explain why
no knots have been associated to FUor burst episodes.

A fundamental quantity in the star formation process is also given by the mass ejection to accretion rate ratio ($\dot{M}_{out}$/$\dot{M}_{acc}$),
which provides us with the efficiency of the stellar accretion process. At variance with the single terms of the ratio, which, on average, depends on $M_*$
and decreases as the source evolves~\citep[][]{natta06,caratti12}, $\dot{M}_{out}$/$\dot{M}_{acc}$ seems to be relatively constant in time (between $\sim$0.01 and $\sim$0.1),
according to both MHD models and observations~\citep[e.\,g.,][]{calvet,pudritz_PPV}. 
It has also been stressed~\citep[e.\,g.,][and references therein]{cabrit09} that such a ratio should not exceed a value of about 0.1, in order to agree 
with the main MHD launching models. 
Indeed, the measured ratios are strongly affected by the uncertainty on both $\dot{M}_{out}$ and $\dot{M}_{acc}$ estimates, 
usually up to one order of magnitude. From our estimates in Sect.~\ref{accretion:sec} and \ref{mjet:sec} we would obtain a ratio very close to one, 
which is too high, even when uncertainties are taken into account.
However, we should keep in mind that the inferred estimates might not trace two simultaneous events. 
In particular, due to the different sizes of the emitting regions, the different cooling times of the two considered species, and 
the low spatial resolution of our spectral-images (0\farcs8$\sim$400\,AU),
the ejection rate derived from the [\ion{Fe}{ii}] emission is averaged over $\sim$6 years around the outburst event, whereas the \ion{H}{i} emission is mostly
tracing the present accretion rate. Therefore we should instead compare $\dot{M}_{out}$ with an average value of $\dot{M}_{acc}$ during and after the 
outburst~\citep[few 10$^{-6}$\,M$_{\sun}$\,yr$^{-1}$;][]{lorenzetti09,kun}. The resulting ratio is then between 0.05 and 0.1.
A similar result is obtained comparing the actual $\dot{M}_{out}$/$\dot{M}_{acc}$ ratio, derived from the \ion{H}{i} in Sect.~\ref{accretion:sec} and
Sect.~\ref{HImodel:sec}.

% how close is PV cep to other Exors?

%-Comparison with literature data

%$\dot{M}_{out}$/$\dot{M}_{acc}$ ratio on source is not correct in case of outburst (low spatial resolution): le dimensioni delle regioni HI e FeII sono estremamente
%differenti: $\dot{M}_{acc}$ misura il valore attuale, mentre $\dot{M}_{out}$ e' mediato lungo il jet: in caso di outburst nel passato recente
%$\dot{M}_{out}$/$\dot{M}_{acc}$ e' molto grande, mentre se e' in corso e' piu' basso. 

\subsection{An ionised jet/wind}
\label{wind:sec}

The fading phase of PV\,Cep gives us a unique opportunity to study the jet and outflow relatively close to the source,
also providing us with indications of the accretion/ejection interplay.

Our analysis shows strong evidence of both jet and wind in the PV\,Cep outflowing material. 
High-resolution spectroscopy of YSO jets usually reveals the existence of two velocity components close to the source, namely a high- and low-velocity
component. The HVC is associated with the extended collimated jet. On the other hand, the LVC is usually confined close to the central 
object (up to 100--200\,AU) and it is probably produced by a disc wind ~\citep[see e.\,g.,][]{rebeca08,dougados08,caratti09}. 
Both components are usually observed through forbidden emission lines (FELs) in CTTSs and in a few Herbig Ae/Be stars~\citep[see, e.\,g.,][]{pyo02,davis03},
whereas molecular hydrogen emission lines (MHELs) have been also detected in less evolved Class\,I objects~\citep[see, e.\,g.,][]{davis_MHEL}. 
Moreover, the LVC is confined within $\sim$200\,AU in CTTSs, whereas in the younger and less evolved Class\,I objects is usually more extended~\citep[up to
$\sim$1000--2000\,AU; see][]{rebeca08,rebeca10}.
The former picture is more likely in PV\,Cep, in which the [\ion{Fe}{ii}] LV component is positioned close to the source ($\leq$400\,AU), 
whereas the HV component is extended. 
In addition, the LVC (in both lobes) has a wide $FWHM$ ($\Delta v\sim$160\,km\,s$^{-1}$, see Sect.~\ref{jetkinematics:sec}), 
which is 2--3 times larger than the HVC $FWHM$ ($\Delta v\sim$70\,km\,s$^{-1}$).
Because MHD models predict that the jet terminal velocity is mostly determined by the Alfv\'en radius, within which the magnetic field
lines act as lever arms~\citep[see e.\,g.,][]{konigl00}, the width of the line is closely related to the size of the emitting region~\citep[][]{pyo02}.
Thus the narrow velocity width of the HVC indicates that the line-emitting jet is launched within a narrow region in the disc, whereas the broad velocity
width of the LVC suggests a wider region of the disc. Moreover, the higher peak velocity of the HVC implies that it is launched from a disc region more internal
than that of the LVC. This indicates that the HV and LV components trace the jet and the wind, respectively.
Notably, the described picture fits well the CO observations in \citet{arce}. 
Based on their analysis of the geometry and kinematics of the molecular emission close to the source, the authors predicted the presence of a wide-angle wind
and a collimated jet.
Finally, we also note that the jet is fully atomic. Indeed the H$_2$ line at 2.12\,$\mu$m (the only H$_2$ line detected in the spectrum) 
is not produced along the jet, which is fast enough ($\sim$600\,km\,s$^{-1}$) to dissociate the H$_2$ and partially ionise the medium. 
At variance with the [\ion{Fe}{ii}] emission, the 1-0\,S(3) line is not spatially resolved and it is confined within the first 400\,AU from the source. 
It shows just one velocity component at $\sim$$-$13($\pm$15)\,km\,s$^{-1}$,
i.\,e., very close to the rest velocity frame of the system, with a narrow $FWHM$ ($\sim$7.4\,\AA, i.\,e. $\Delta v \sim$60\,km\,s$^{-1}$), 
pointing to a disc and/or wind origin. The former is more likely, due to the lack of a red-shifted component.

\subsection{Origin of the jet/wind asymmetry}
\label{asymmetry:sec}

Another interesting and peculiar aspect of the PV\,Cep jet/outflow is its asymmetry. Velocities in the southern red-shifted and northern blue-shifted
lobes are asymmetric in both LV and HV components, as well as their inferred electron densities. Asymmetric jets from YSOs have already been observed
in various YSOs~\citep[see e.\,g.,][]{hirth94,melnikov,podio11}, but {\it this is the first documented case of an asymmetric jet produced during an outburst.}

The origin of such asymmetries still remains an open problem. It might
be intrinsic to the source and strictly linked to the launching mechanism (e.\,g., asymmetry in the magnetic field configuration), or extrinsic, i.\,e. caused
by different physical conditions in the circumstellar medium, such as an inhomogeneous medium~\citep[for a detailed discussion see,][]{matsakos}.
In principle, the former scenario could explain both the observed YSO variability (which cannot be triggered by the presence of a close companion)
and the asymmetries observed in the jet. 

However, the estimated mass loss rate in the two lobes is comparable, suggesting that the ejection power is similar 
on both sides of the system, as expected from a magneto-centrifugal ejection mechanism. Notably, two previously 
analysed asymmetric jets~\citep[namely, \object{RW Aur} and \object{DG Tau}; see][]{melnikov,podio11} show similar characteristics: velocity
and density asymmetries and the same $\dot{M}_{out}$ in both lobes. The authors of both works suggest that
the observed asymmetries are due to different mass load and/or
propagation properties in an inhomogeneous environment. 
The medium around PV\,Cep is highly inhomogeneous, as observed in the large-scale CO outflow studied by \citet{arce02}, and partially suggested 
by the different $A_{\rm V}$ values observed along the jet in the northern (blue-shifted) and southern (red-shifted) lobes (see Sect.~\ref{reddening:sec}).
PV\,Cep is located in the northern edge of its dark cloud, thus the northern and southern lobes interact with environments at very different densities.
Hence, the blue-shifted ejected matter (northwards) interacts with a much less dense medium, producing a highly asymmetric jet and, in turn, a highly asymmetric outflow.
It is worth noting that the velocity asymmetry in the [\ion{Fe}{ii}] LV component implies that such inhomogeneities in the medium density must extend very close
to the YSO (at least down to 400\,AU), and, more likely, down to the YSO disc. In this case, the disc would directly interact with an inhomogeneous asymmetric 
ambient medium, producing an asymmetry in magnetic lever arms and/or launch radii between either sides of the disc~\citep[][]{ferreira06,podio11}.
Thus, the power generated by the MHD dynamo and conveyed by the rotating disc to both sides is identical, but the physical conditions of the ejected material
may be different on the two sides.

%-Mout the bulk of the mass is transported by the HVC, whereas LVC is confined close to the source (disc wind? --> larger FWHM).

\subsection{Circumstellar environment}
\label{circumstellar:sec}

\begin{figure}
\centering
\includegraphics [width=9cm]{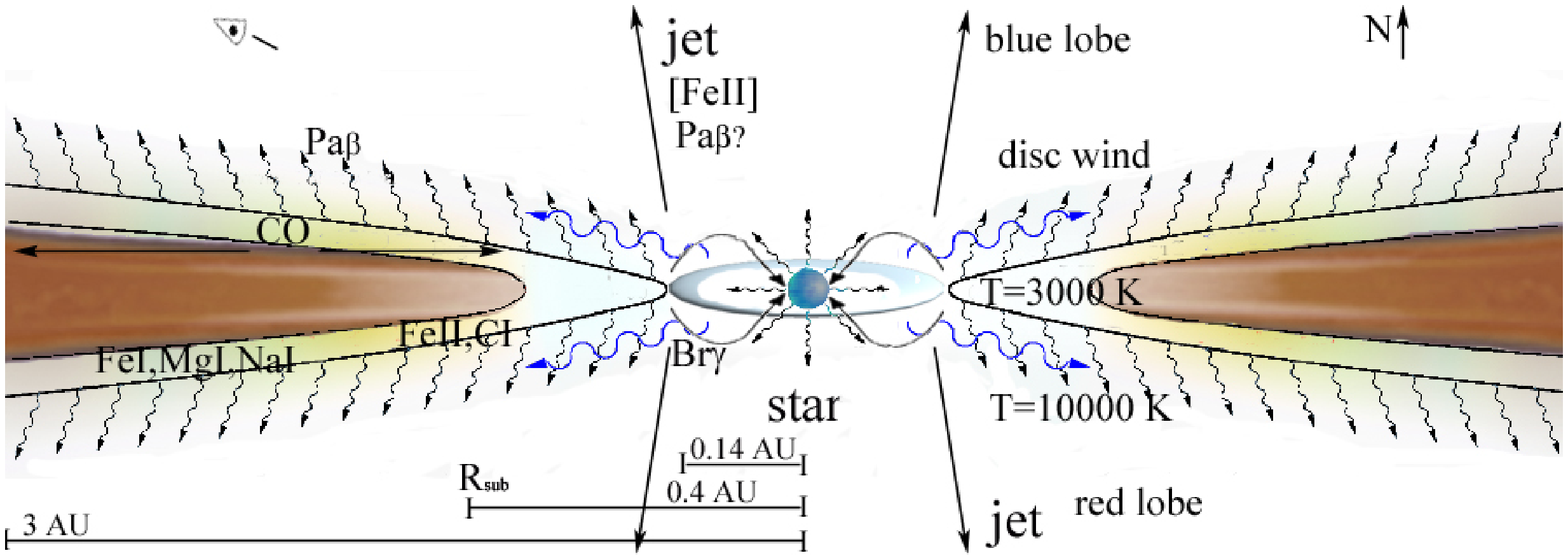}\\
\caption{Sketch (not in scale) of PV\,Cep's innermost circumstellar region.}
\label{fig:CSM}
\end{figure}

Our analysis provides us with an overview of the circumstellar environment of PV\,Cep during the pre-quiescent phase,
i.\,e. when the outbursting phase is fading and the inner disc region is rearranging in a more stable state.
The inner region is dominated by hydrogen gas (up to $\sim$0.14\,AU) as delineated by the Br$\gamma$ line (see Fig.\ref{fig:CSM}). 
A more extended \ion{H}{i} region (up to few AUs) is also traced by the Paschen emission lines. 
In this region there is also evidence of atomic fluorescent emission lines (\ion{Fe}{ii} at 1 and 1.688\,$\mu$m; 
and \ion{C}{i} at 1.17--1.19\,$\mu$m) and other metallic lines (e.\,g., \ion{Fe}{i}, \ion{Fe}{ii}, \ion{Mg}{i}, \ion{C}{i}, \ion{Na}{i}),
whose excitation potential (4--8\,eV) is lower than \ion{H}{i} ($\sim$13.6\,eV). According to \citet{kospal11}, 
the emitting region is placed where the hydrogen is mainly neutral, likely shielded from the shock radiation by the circumstellar disc.
On the other hand, the fluorescent emission indicates that the emitting region is located close to the disc surface.
Beyond this region, we model a small-scale gaseous disc ring, traced by the CO emission, and extending from $\sim$0.2-0.4\,AU to $\sim$3\,AU from the source.
The inner region has a temperature of $\sim$3000\,K. This part of the disc is hotter in comparison with the typical values found in Herbig Ae stars. 
This might be a result of the heat produced by the outburst.
Dust is likely to coexist within this gaseous region, $R_{sub}$ being at 0.3--0.5\,AU from the source.

Finally, [\ion{Fe}{ii}] analysis proves the existence of both wind and jet emission close to the source.
Our modelling suggests that the bulk of the \ion{H}{i} emission originates from a wind located close to the source.
The presence of a secondary maximum in the Paschen lines also indicates that a small part of the Paschen emission might
originate from the blue-shifted part of the jet.

%%%%%%%%%%%%%%%%%%%%%%%%%%%%%%	Fe II profile  %%%%%%%%%%%%%%%%%%%%%%%%%%%%%%%%%%%%%%%%%%%%%%%%%%%%%%%%%%%%%%%%%%%%%%%%%%%
%%%%%%%%%%%%%%%%%%%%%%%%%%%%%%

\subsection{How close is PV\,Cep to EXors?}
\label{exor:sec}

An interesting question to answer is how similar PV\,Cep is to other EXors. 
\citet{lorenzetti11} already pointed out that PV\,Cep is not a genuine EXor according to the classical definition given by \citet{herbig89}.
As considered by the authors, the main resemblance with the EXor group resides 
in both spectrum and recurrent outbursts driven by disc accretion episodes. On the other hand, 
the most important discrepancies appear to be its higher mass and earlier evolutionary stage with respect to other classical EXors.
The remaining differences listed in \citet{lorenzetti11} (i.\,e. circumstellar nebulosity, jet/outflow presence, massive disc, far-infrared excess
in the SED, presence of maser and radio continuum emission) can all be a consequence of the previous two.

We can now compare our results with the circumstellar environment of the EXor prototype EX\,Lup~\citep[][]{kospal11,sicilia} to get a more quantitative evaluation.
Our analysis shows many similarities between the two objects, both in the physical structure 
and the observed spectra. However, as already mentioned in Sect.~\ref{COmodel:sec} and ~\ref{HImodel:sec}, the size of the
various regions appears to be larger and the excitation conditions are more extreme. This is 
compatible with a  hotter, more massive object. 
Another significant difference in PV\,Cep spectrum is the presence of jet tracers.
Conical nebulae (namely the outflow cavities), jets, knots and outflows are not present in EXors.
This indicates that the PV\,Cep is young and that its circumstellar environment has not been cleared by the YSO activity, favouring the formation
of a jet, shocked knots, and a swept out outflow.

In conclusion, as already pointed out by \citet{lorenzetti11}, PV\,Cep is not a classical EXor because it is not a low-mass PMS star.
It is more massive, young and embedded than classical EXors. Nevertheless, it shows the signature of an `EXor-like' outburst,
which suggests a common origin. We can therefore argue that PV\,Cep is a younger, high-mass counterpart of classical EXors.
Indeed, the discovery and study of more embedded outbursting YSOs in the near future will increase our knowledge of and statistics on these phenomena 
and possibly confirm that enhanced accretion processes are common to all YSOs, 
despite the difference in their mass or evolutionary stage. 
Therefore, on the basis of the new discoveries, it is quite likely that the two main sub-classes (FUors and EXors) 
will be revised.

%How does the outburst affected the circumstellar environment?

\section{Conclusions} 
\label{conclusion:sec}

We have presented near-IR medium-resolution spectroscopy of PV\,Cep taken with LUCIFER at the LBT in 2012, i.\,e.
eight years after its last outburst. By means of the many emission lines detected in the spectrum, we are able to 
trace both accretion and ejection activity, deriving the physical and dynamical conditions in its circumstellar
region. 
The main results of this work are the following:
\begin{enumerate}
\item[-] The NIR spectrum of PV\,Cep displays several strong emission lines and a steeply rising continuum without photospheric absorption features.
The most prominent features in the PV\,Cep spectrum are circumstellar features, originating from YSO accretion or inner winds, as, e.\,g., 
\ion{H}{i}, \ion{Ca}{ii}, shocks along the jet ([\ion{Fe}{ii}]), inner disc region (e.\,g.,\ion{Na}{i}, CO) or chromospheric activity 
(e.\,g., \ion{Fe}{i}, \ion{Fe}{ii}, \ion{Mg}{i}, \ion{C}{i}). A few fluorescent emission lines from \ion{Fe}{ii} (i.\,e., at 1 and 1.688\,$\mu$m) 
and \ion{C}{i} (i.\,e., 1.17--1.19\,$\mu$m) are pumped by UV photons.
\item[-] Compared to the outburst in 2004 and subsequent photometry, our data show that PV\,Cep's brightness is fading, indicating that the outburst phase is declining.
However, the many detected lines show that accretion and ejection activities still proceed at high rate.
\item[-] The actual mass accretion rate is 2$\times$10$^{-7}$\,M$_{\sun}$\,yr$^{-1}$, i.\,e. at least more than one order of magnitude lower than values measured
during the 2004 outburst. 
%The current value is still one order of magnitude higher than typical Herbig Ae accretion rates: PV\,Cep
%is younger and still accreting at higher rates, as is also expected from the presence of a massive disc.
\item[-] Among the several emission lines, only the [\ion{Fe}{ii}] intensity increased since the outburst in 2004. The observed emission traces blue- and red-shifted lobes
of an asymmetric jet/outflow. In each lobe, two velocity components are detected: a LVC, not spatially resolved (within 400\,AU from the source), likely tracing
a disc wind, and an extended HVC tracing the jet. Velocities between the two lobes are asymmetric ($\sim$1.6 ratio between blue- and red-shifted part). 
Total velocities for the HVC and LVC are $-$570 and 350\,km\,s$^{-1}$, and $-$210 and 130\,km\,s$^{-1}$, respectively.
\item[-] The observed emission has a dynamical age of 7--8 years, indicating that it was produced during the 2004 outburst. 
\item[-] Jet physical properties are derived from different [\ion{Fe}{ii}] transition ratios. The visual extinction decreases moving from 
the red-shifted ($A_{\rm V}(red)$=10.1$\pm$0.7\,mag) to the blue-shifted lobe ($A_{\rm V}(blue)$=6.5$\pm$0.4\,mag). We derive an average electron 
temperature of $T_{\rm e}$=17\,500\,K, and electron densities of 30\,000\,cm$^{-3}$ and 15\,000\,cm$^{-3}$ for the blue and the red lobe, respectively.
\item[-] The measured mass ejection rate is $\sim$1.5$\times$10$^{-7}$\,M$_{\sun}$\,yr$^{-1}$ in both lobes. 
These estimates fairly match the high accretion rate observed during and immediately after 
the outburst ($\dot{M}_{out}$/$\dot{M}_{acc}$$\sim$0.05--0.1).
\item[-] The similar mass loss rate in the two lobes suggests that the ejection power is comparable on both sides of the disc,
as expected from MHD ejection mechanisms.
The observed asymmetries are thus consistent with an inhomogeneous medium, as also indicated by the large-scale CO analysis in \citet{arce}.
\item[-] Our modelling of the CO emission hints at a small-scale gaseous disc ring, extending from $\sim$0.2-0.4\,AU to $\sim$3\,AU from the source. This
region has an inner temperature of $\sim$3000\,K (exponentially decreasing).
\item[-] Our \ion{H}{i} line modelling indicates that most of the observed emission comes from an expanding wind at $T_{\rm e}$=10\,000\,K. Brackett and
Paschen line profiles are strongly affected by scattering, disc screening, and outflow self-absorption.
\item[-] PV\,Cep is not an EXor object, according to the classical definition given by \citet{herbig89}. It is more massive and younger than typical EXors.
Nevertheless, it shows the signature of an `EXor-like' outburst, suggesting a similar outburst mechanism.

In conclusion, EXors and related objects offer a unique opportunity to investigate the fundamental 
accretion/ejection mechanism, because the physical properties of the accretion process and, only few years later, the
corresponding ejection event can be related and studied in detail.

\end{enumerate}

\begin{acknowledgements}
We wish to thank the referee, Dr. \'Agnes K{\'o}sp{\'a}l, for her useful insights and comments.
J.D. Ilee acknowledges funding from the European Union FP7$-$2011 under grant agreement no. 284405.
This research has also made use of NASA's Astrophysics Data System Bibliographic Services and the SIMBAD database, operated
at the CDS, Strasbourg, France.
\end{acknowledgements}

\bibliographystyle{aa}
\bibliography{references}

\Online
\begin{appendix}

%%%%%%%%%%%%%%%%%%%%%%%% TAB 1 %%%%%%%%%%%%%%%%
\longtab{1}{
\begin{longtable}{ccccccccc}
\caption{\label{lines:tab} Observed emission lines in the spectrum of PV\,Cephei}\\
\hline \hline\\[-5pt] 
$\lambda_{obs}$ &  F & $\Delta$F & W$_\lambda$  & FWHM & S/N & ID (term) & $\lambda_{vac}$ &  $v_{\rm r}\,(peak)$ \\
   (\AA)      &  \multicolumn{2}{c}{(10$^{-15}$\,erg\,cm$^{-2}$\,s$^{-1}$)} & (\AA) &  (\AA) & ratio & & (\AA) & (km\,s$^{-1}$) \\
\hline\\[-5pt]
\endfirsthead
\caption{continued.}\\
\hline\hline\\[-5pt] 
$\lambda_{obs}$ &  F & $\Delta$F & W$_\lambda$  & FWHM & S/N & ID (term) & $\lambda_{vac}$ &   $ v_{\rm r}(peak)$ \\
   (\AA)      &  \multicolumn{2}{c}{(10$^{-15}$\,erg\,cm$^{-2}$\,s$^{-1}$)} & (\AA) &  (\AA) & ratio & & (\AA) & (km\,s$^{-1}$) \\
\hline\\[-5pt]
\endhead
\hline
\endfoot
9173.9  & 11.4  & 2.1 & $-$10.4  & 6.4 &  5    &  \ion{Fe}{ii}  &  9175.2  &  $-$18       \\ 
9230.4  & 6.0	& 1.0 & $-$6.2   & 5.3 &  6    &  Pa\,9    &  9231.6  &  $-$14      \\ 
9261.3  & 2.3	& 0.5 & $-$2.1   & 3.1 &  5    &  [\ion{Fe}{ii}] ($a^4\!P_{1/2}-a^4\!F_{3/2}$) &  9270.1  &  $-$262     \\ 
9994.4  & 1.7	& 0.5 & $-$1.1   & 4.3 &  3    &  \ion{Fe}{i} ?  &  9994.0  &  38      \\ 
10000.3 & 3.2  & 0.7 &  $-$2.1   & 3.8 &  5   &   \ion{Fe}{ii}     &  10000.3 &  20       \\ 
10042.7 & 5.8  & 0.9 & $-$1.6   & 7.2 &  6   &  Pa$\delta$    &  10052.2 &  $-$259       \\ 
10052.4 & 15.6  & 1.0 & $-$8.7   & 7.7 &  16   &  Pa$\delta$    &  10052.2 &  31       \\ 
11640.5 & 3.2	& 0.3 & $-$1.1   & 4.3 &  11   &  \ion{Fe}{i}    &  11641.5 &  0        \\ 
11661.6 & 2.0	& 0.3 & $-$0.8   & 4.2 &  7    &  \ion{C}{i}   &  11662.9 &  $-$8        \\ 
11673.6 & 2.3	& 0.5 & $-$0.8   & 6.6 &  5    &  \ion{C}{i} + \ion{Fe}{i} ?  &  11672.8 &  44        \\ 
11692.2 & 4.9	& 0.4 & $-$1.8   & 5.0 &  14   &  \ion{Fe}{i}    &  11693.2 &  0        \\ 
11750.4 & 4.0	& 0.4 & $-$1.4   & 6.2 &  9    &  \ion{C}{i}     &  11751.5 &  $-$2      \\ 
11756.2 & 13.0  & 0.5 & $-$4.4   & 7.1 &  26   &  \ion{C}{i}     &  11756.6 &  14       \\ 
11780.0 & 1.7	& 0.3 & $-$0.6   & 4.0 &  6    &  \ion{C}{i}     &  11780.8 &  5        \\ 
11785.4 & 5.5	& 0.4 & $-$1.9   & 6.1 &  13   &  \ion{Fe}{i}    &  11786.5 &  $-$3       \\ 
11830.2 & 13.1  & 0.5 & $-$4.3   & 6.5 &  29   &  \ion{Mg}{i}    &  11831.4 &  $-$6       \\ 
11840.9 & 10.7  & 0.5 & $-$3.5   & 6.9 &  22   &  \ion{Ca}{ii}   &  11842.2 &  $-$9       \\ 
11885.1 & 14.8  & 0.5 & $-$4.7   & 5.7 &  32   &  \ion{P}{ii} +   &  11886.1 &  $-$1       \\ 
11885.1 &       &     &         &    &       & \ion{Fe}{i}     &  11886.1 &  $-$1       \\ 
11894.9 & 1.5	& 0.4 & $-$0.7   & 6.3 &  4    &  \ion{C}{i}     &  11896.2 &  $-$7       \\ 
11897.5 & 2.1	& 0.4 & $-$1.1   & 6.7 &  5    &  \ion{C}{i}     &  11899.0 &  $-$13      \\ 
11952.2 & 6.9	& 0.5 & $-$2.1   & 5.9 &  13   &  \ion{Ca}{ii}   &  11953.0 &  4        \\ 
11975.1 & 13.9  & 0.5 & $-$4.4   & 5.8 &  27   &  \ion{Fe}{i}    &  11976.3 &  $-$6       \\ 
11986.0 & 8.8	& 0.5 & $-$2.8   & 5.5 &  18   &  \ion{Ca}{ii} + &  11987.5 &  $-$13      \\ 
11986.0 &       &     &       &       &      &  \ion{Si}{i}   &  11987.5 &  $-$13  \\ 
11993.5 & 6.2	& 0.5 & $-$2.0   & 5.3 &  11   &  \ion{Ca}{ii} + &  11994.8 &  $-$9       \\ 
11993.5 &  	&     &        &     &       &  \ion{Si}{i}   &  11994.8 &  $-$9        \\ 
12033.6 & 12.1  & 0.6 & $-$4.1   & 6.3 &  19   &  \ion{Si}{i}   &  12034.8 &  $-$5        \\ 
12085.9 & 5.2	& 0.6 & $-$1.6   & 5.8 &  9    &  \ion{Mg}{i}    &  12086.6 &  8        \\ 
12105.7 & 4.2	& 0.5 & $-$1.3   & 5.2 &  8    &  \ion{Si}{i}   &  12106.9 &  $-$4        \\ 
12272.9 & 6.4	& 0.4 & $-$1.8   & 6.7 &  16   &  \ion{Si}{i}   &  12274.1 &  $-$3        \\ 
12558.2 & 16.6  & 0.2 & $-$4.4   & 4.0 &  67   &  [\ion{Fe}{ii}] ($a^4\!D_{7/2}-a^6\!D_{9/2}$) &  12570.2 &  $-$263     \\ 
12565.2 & 9.9	& 0.3 & $-$3.0   & 6.8 &  23   &  [\ion{Fe}{ii}] ($a^4\!D_{7/2}-a^6\!D_{9/2}$) &  12570.2 &  $-$96      \\ 
12570.9 & 3.7	& 0.4 & $-$1.5   & 7.0 &  13   &  [\ion{Fe}{ii}] ($a^4\!D_{7/2}-a^6\!D_{9/2}$) &  12570.2 &  40       \\ 
12575.9 & 7.0	& 0.2 & $-$1.8   & 3.4 &  34   &  [\ion{Fe}{ii}] ($a^4\!D_{7/2}-a^6\!D_{9/2}$) &  12570.2 &  160      \\ 
12616.9 & 2.2	& 0.4 & $-$0.6   & 6.9 &  5    &  \ion{C}{i}     &  12617.6 &  9        \\ 
12694.6 & 2.0	& 0.3 & $-$0.5   & 3.3 &  7    &  \ion{Si}{i} ?  &  12694.2 &  32        \\ 
        &  	&     &        &     &       &  \ion{Al}{i} ?  &  12695.0 &  15        \\ 
12779.0 & 1.9	& 0.3 & $-$0.5   & 3.1 &  7    &  [\ion{Fe}{ii}] ($a^4\!D_{3/2}-a^6\!D_{3/2}$) &  12791.3 &  $-$263     \\ 
12785.9 & 1.1	& 0.2 & $-$0.3   & 4.3 &  5    &  [\ion{Fe}{ii}] ($a^4\!D_{3/2}-a^6\!D_{3/2}$) &  12791.3 &  $-$101     \\ 
12796.8 & 1.7	& 0.5 & $-$0.4   & 5.3 &  4    &  [\ion{Fe}{ii}] ($a^4\!D_{3/2}-a^6\!D_{3/2}$)&  12791.3 &  154      \\ 
12808.7 & 4.4  & 0.2 & $-$1.0  & 4.1 &  20     &  Pa$\beta$    &  12821.7 & $-$279       \\ 
12821.7 & 113.0 & 0.8 & $-$26.8  & 9.4 &  134  &  Pa$\beta$    &  12821.7 &  25       \\ 
12933.8 & 3.4	& 0.5 & $-$1.0   & 4.3 &  6    &  [\ion{Fe}{ii}] ($a^4\!D_{5/2}-a^6\!D_{5/2}$)&  12946.2 &  $-$263     \\ 
13125.6 & 5.5	& 0.6 & $-$1.7   & 5.0 &  9    &  \ion{Al}{i}   &  13127.0 &  $-$7       \\ 
15441.6 & 16.5  & 1.1 & $-$4.0   & 12.3&  15   &  Br\,17 + ? &  15443.2 &  $-$6       \\ 
15559.0 & 24.0  & 1.0 & $-$4.5   & 11.1&  24   &  Br\,16   &  15560.8 &  $-$9       \\
15624.3 & 3.7	& 1.0 & $-$1  & 7.5    & 4     &  ?    &$\cdots$&$\cdots$ \\
15703.1 & 24.0  & 1.0 & $-$4.5   & 10.7&  25   &  Br\,15   &  15705.0 &  $-$12      \\ 
15742.2 & 1.9	& 0.5 & $-$0.6   & 5.4 &  4    &  \ion{Mg}{i}    &  15745.0 &  $-$29      \\ 
15750.2 & 10.5  & 1.0 & $-$3.1   & 11.5&  10   &  \ion{Mg}{i} + ?  &  15753.2 &  $-$32      \\ 
15769.5 & 18.0  & 0.8 & $-$3.0   & 8.6 &  23   &  \ion{Mg}{i} + ?  &  15770.0 &  15       \\ 
15883.0 & 38.0  & 1.3 & $-$7.0   & 11.0&  29   &  Br\,14   &  15885.0 &  $-$12      \\ 
15889.5 & 42.0  & 1.3 & $-$7.0   & 11.0&  32   &  \ion{Mg}{i}    &  15890.6 &  4        \\ 
15961.4 & 3.3	& 0.7 & $-$0.6   & 5.6 &  5    &  \ion{Si}{i} + ? &  15964.0 &  $-$25       \\ 
15965.8 & 3.6	& 0.7 & $-$0.6   & 5.8 &  5    &  \ion{Fe}{i}  &  15966.9 &  4 \\ 
15983.3 & 6.3	& 0.8 & $-$1.0   & 7.0 &  8    &  [\ion{Fe}{ii}] ($a^4\!D_{3/2}-a^4\!F_{7/2}$) &  15999.1 &  $-$272     \\ 
16007.1 & 5.3	& 0.7 & $-$0.9   & 5.9 &  7    &  [\ion{Fe}{ii}] ($a^4\!D_{3/2}-a^4\!F_{7/2}$) &  15999.1 &  175      \\ 
16111.6 & 41.3  & 1.4 & $-$6.7   & 11.6&  30   &  Br\,13   &  16113.8 &  $-$16      \\ 
16382.3 & 5.4	& 0.8 & $-$0.8   & 6.5 &  7    &  \ion{Si}{i} + \ion{Fe}{i}   &  16382.7 &  15       \\ 
16409.3 & 55.1  & 1.5 & $-$8.4   & 9.3 &  30   &  Br\,12   &  16411.7 &  $-$20      \\ 
16424.1 & 21.1  & 0.8 & $-$3.1   & 6.0 &  27   &  [\ion{Fe}{ii}] ($a^4\!D_{7/2}-a^4\!F_{9/2}$) &  16440.0 &  $-$265     \\ 
16433.2 & 14.6	& 0.9 & $-$2.1   & 10.0 &  8    &  [\ion{Fe}{ii}] ($a^4\!D_{7/2}-a^4\!F_{9/2}$) &  16440.0 &  $-$99      \\ 
16442.0 & 7.5	& 0.6 & $-$1.1   & 10.0 &  5    &  [\ion{Fe}{ii}] ($a^4\!D_{7/2}-a^4\!F_{9/2}$) &  16440.0 &  61       \\ 
16447.7 & 13.4  & 0.5 & $-$2.4   & 4.5 &  28   &  [\ion{Fe}{ii}] ($a^4\!D_{7/2}-a^4\!F_{9/2}$) &  16440.0 &  165      \\ 
16626.0 & 2.4	& 0.6 & $-$0.3   & 4.8 &  4    &  [\ion{Fe}{ii}] ($a^4\!D_{1/2}-a^4\!F_{5/2}$) &  16642.2 &  $-$267     \\ 
16757.3 & 4.8	& 0.6 & $-$0.7   & 4.8 &  8    &  [\ion{Fe}{ii}] ($a^4\!D_{5/2}-a^4\!F_{7/2}$) &  16773.3 &  $-$262     \\ 
16781.3 & 4.3	& 0.8 & $-$0.7   & 7.0 &  5    &  [\ion{Fe}{ii}] ($a^4\!D_{5/2}-a^4\!F_{7/2}$) &  16773.3 &  167      \\ 
16808.9 & 62.0  & 1.4 & $-$8.4   & 11.4&  45   &  Br\,11   &  16811.2 &  $-$16      \\ 
16875.8 & 6.2  & 1.3 & $-$0.9   & 8.5 &  5   &  \ion{Fe}{ii}     &  16877.8 &  $-$11       \\ 
16893.4 & 18.3  & 1.3 & $-$2.5   & 9.0 &  15   &  \ion{C}{i}     &  16895.0 &  $-$4       \\ 
17111.7 & 25.4  & 1.4 & $-$3.5   & 9.8 &  19   &  \ion{Mg}{i}    &  17113.3 &  $-$4       \\ 
17327.6 & 5.4	& 1.2 & $-$0.9   & 8.6 &  5    &  ? &$\cdots$&$\cdots$\\ 
17337.5 & 4.4	& 1.6 & $-$0.7   & 11.5&  3    & \ion{Fe}{ii} &$\cdots$&$\cdots$        \\ 
17344.5 & 6.0	& 1.9 & $-$1.2   & 13.5&  3    &  \ion{Na}{i}  &  17345.8 &  2  \\ 
17364.3 & 85.4  & 1.8 & $-$10.6  & 12.8&  48   &  Br\,10   &  17366.9 &  $-$21      \\ 
17406.5 & 12.0  & 1.2 & $-$1.7   & 8.7 &  10   &  \ion{Fe}{i} + ? &  17406.9 &  17        \\ 
21215.6 & 12.6  & 1.8 & $-$1.5   & 7.4 &  9    &  H$_2$ (1--0\,S(1))   &  21218.3 &  $-$13      \\ 
21658.1 & 193.0 & 3.0 & $-$16.6  & 15.5&  66   &  Br$\gamma$    &  21661.3 &  $-$20      \\ 
22061.0 & 25.6  & 2.2 & $-$2.1   & 11.2&  11   &  \ion{Na}{i}    &  22062.4 &  6        \\ 
22087.4 & 21.8  & 1.6 & $-$1.8   & 8.2 &  13   &  \ion{Na}{i}    &  22089.7 &  $-$7       \\ 
22711.8 & 18.8  & 3.6 & $-$1.5   & 16.2&  5    &  ? &$\cdots$&$\cdots$  \\ 
22937.2 & 267.0 & 3.6 &        &     &  5    &  CO\,(2-0)     &  22935.0 &  53       \\ 
23235.0 & 200.0 & 3.6 & $-$19.0  & 41.0&  5    &  CO\,(3-1)     &  23227.0 &  128      \\  

\hline
\end{longtable}
}
%%%%%%%%%%%%%%%%%%%%%%%%%%%%%%%%%%%%%%%%%%%%%%%%%%%%%%%%%%%
\end{appendix}
\end{document}